\documentclass{sig-alternate-05-2015}

\usepackage{booktabs} 
\usepackage{color}
\usepackage{textcomp}
\usepackage{booktabs}
\usepackage{graphicx}
\usepackage{todonotes}
\usepackage{multirow}
\usepackage{adjustbox}

\begin{document}
\title{Influence of Reviewer Interaction Network on Long-term Citations: A Case Study of the \\ Scientific Peer-Review System of the Journal of High Energy Physics}

\numberofauthors{4} 
%
 \author{
\alignauthor
Sandipan Sikdar\\
       \affaddr{Dept. of CSE}\\
       \affaddr{IIT Kharagpur}\\
       \affaddr{West Bengal, India -- 721302}\\
       \email{sandipansikdar@cse\\.iitkgp.ernet.in}
\alignauthor
Matteo Marsili\\
       \affaddr{ICTP}\\
       \affaddr{Strada Costiera}\\
       \affaddr{34014 Trieste, Italy}\\
       \email{marsili@ictp.it}
 \and
\alignauthor Niloy Ganguly\\
       \affaddr{Dept. of CSE}\\
       \affaddr{IIT Kharagpur}\\
       \affaddr{West Bengal, India -- 721302}\\
       \email{niloy@cse.iitkgp.ernet.in}
\alignauthor Animesh Mukherjee\\
       \affaddr{Dept. of CSE}\\
       \affaddr{IIT Kharagpur}\\
       \affaddr{West Bengal, India -- 721302}\\
       \email{animeshm@cse.iitkgp.ernet.in}
 }

\maketitle
\begin{abstract}
A `peer-review system' in the context of judging research contributions, is one of the prime steps undertaken to ensure the quality 
of the submissions received; a significant portion of the publishing budget is spent towards successful completion of the peer-review by the publication houses. Nevertheless, the scientific community is largely reaching a consensus that peer-review system, although indispensable, is nonetheless flawed.
A very pertinent question therefore is ``could this system be improved?". 
In this paper, we attempt to present an answer to this question by considering a massive dataset of around $29k$ papers with roughly $70k$ distinct review reports together 
consisting of $12m$ lines of review text from the Journal of High Energy Physics (JHEP) between 1997 and 2015. In specific, we introduce a novel \textit{reviewer-reviewer interaction network} (an edge exists between two reviewers if they were assigned by the same editor) and show that surprisingly the simple structural properties of this network such as degree, clustering coefficient, centrality (closeness, betweenness etc.) serve as strong predictors of the long-term citations (i.e., the overall scientific impact) of a submitted paper. These features, when plugged in a regression model, alone achieves a high $R^2$ of \textbf{0.79} and a low $RMSE$ of \textbf{0.496} in predicting the long-term citations. In addition, we also design a set of supporting features built from the basic characteristics of the submitted papers, the authors and the referees (e.g., the popularity of the submitting author, the acceptance rate history of a referee, the linguistic properties laden in the text of the review reports etc.), which further results in overall improvement with $R^2$ of \textbf{0.81} and $RMSE$ of \textbf{0.46}. Analysis of feature importance shows that the network features constitute the best predictors for this task.   
Although we do not claim to provide a full-fledged reviewer recommendation system (that could potentially replace an editor), our method could be extremely useful in assisting the editors in deciding the acceptance or rejection of a paper, thereby, improving the effectiveness of the peer-review system. 

\end{abstract}



\section{Introduction}

Peer-review system has been relied upon by the scientific community for determining the correctness and the quality of the findings presented in a research article. The authenticity and, hence, the need for this process has long been debated since in many cases flawed research has got into the literature even though the peer-review process was rigorous \cite{bohannon2013s}. Similarly, there have been cases where excellent research was misjudged by the peer-review process and therefore rejected \cite{braatz2014papers}. The publishing house makes significant investments into ensuring the quality of editing and reviewing of the received submissions and, therefore, identifying the necessity of this entire system is of prime importance. 

\noindent{\bf Debates on the scientific peer-review:} 
The effectiveness of peer-review have been studied to a large extent in the domain of medical sciences where peer-review is heavily relied upon for judging the quality of a research article ~\cite{jefferson2006editorial,kassirer1994peer,
rennie1990editorial}. The effect of blinding on the quality of peer review has also been studied in detail~\cite{jefferson2002measuring, mcnutt1990effects}. It was observed that blinding improves the quality of reviews. Several limitations of the review process have also been pointed out~\cite{horrobin1990philosophical}. In~\cite{cole1981chance} the authors show that there is a high degree of disagreement within the population of eligible reviewers.~\cite{braatz2014papers} also shows that there are a significant number of papers that receive more citations after rejection. All these together point to limitations of the review process and have resulted in the scientific community questioning the requirement of this process. 

\noindent{\bf A massive peer-review dataset:} In this paper, we investigate the effectiveness of the peer-review system through a rigorous and large-scale analysis of the scientific review data. In particular, we consider a set of around $29k$ papers along with roughly $70k$ unique review reports containing $12m$ lines of review text submitted to the Journal of High Energy Physics (JHEP) between 1997 and 2015. We would like to point out here that this dataset is unique as well as very rich and we do not know of any other work that presents such a large-scale analytics of an equivalent dataset. Informed with the details of the number of reviews per paper, the content of the review reports and the citation counts we perform, for the first time, a series of systematic measurements to determine whether the peer-review process is indeed able to correctly differentiate between high impact contributions and the rest.

\noindent{\bf Citation impact of accepted papers:}  Assuming that citation count of a paper is representative of its overall quality, we observe that on average those papers which were accepted at JHEP after passing through the peer-review process, are cited more often compared to those which got rejected at JHEP and eventually got accepted at a different venue. While this is true for the majority, there are a few exception cases where either a rejected paper is found to receive high citations or an accepted paper is found to receive (almost) no citation. 

\noindent{\bf Reviewer-reviewer interaction network:} One of the central contributions of this work is the introduction of a novel reviewer-reviewer interaction network built as the one-mode projection of the editor-reviewer bipartite network. The reviewer-reviewer interaction network has nodes as the reviewers and two reviewers are connected by an edge if they have been assigned by the same editor. Surprisingly, the network related structural features such as the degree, the clustering coefficient and the centrality values (closeness, betweenness etc.) of the reviewer nodes in the reviewer-reviewer network strongly correlate with the  long-term citations received by the papers these reviewers refereed. 

\noindent{\bf Supporting features:} Another unique contribution of this paper is that we also build a set of supporting features based on the various characteristics of the papers submitted as well as the authors and the referees of the submitted papers. {\em Papers}: The highly cited papers tend to undergo lesser rounds of review and there also exists an optimal team size (number of contributing authors) for which the accrued citation is maximum.
{\em Review Reports}: For the accepted papers, the length of the review reports seem to be indicator of the long-term citation. Moreover there exists an optimal length for which the citation obtained by the corresponding paper is maximum. On performing sentiment analysis, we observe the review reports to be mostly neutral as the referees hardly use highly polar words in their reports. However, we observe several linguistic quality indicators which can be extracted from the review text that determines whether a paper is going to be cited well in the future. {\em Authors}: From the author specific analysis we observe that for authors who have a higher acceptance to submission ratio tend to receive more citations than others who have a lower acceptance to submission ratio. In addition, the reviews received by authors having higher acceptance to submission ratio tend to contain more positive sentiments on average. {\em Reviewers}: In a previous work~\cite{sikdar2016anomalies} the authors showed that the reviewers who tend to accept or reject most of the papers assigned to them fail to correctly judge the quality of the papers. We include such history based features of the reviewers in the set of supporting features. 

Two further interesting observations from the analysis of the supporting features are -- the low cited accepted papers got in due to the higher accept history of the authors and the lenience of the referees; the high cited rejected papers could not make a place because of lower accept history of the authors and the strictness of the referees.

\begin{figure*}
\centering
\begin{tabular}{ccc}
\includegraphics[scale=.15]{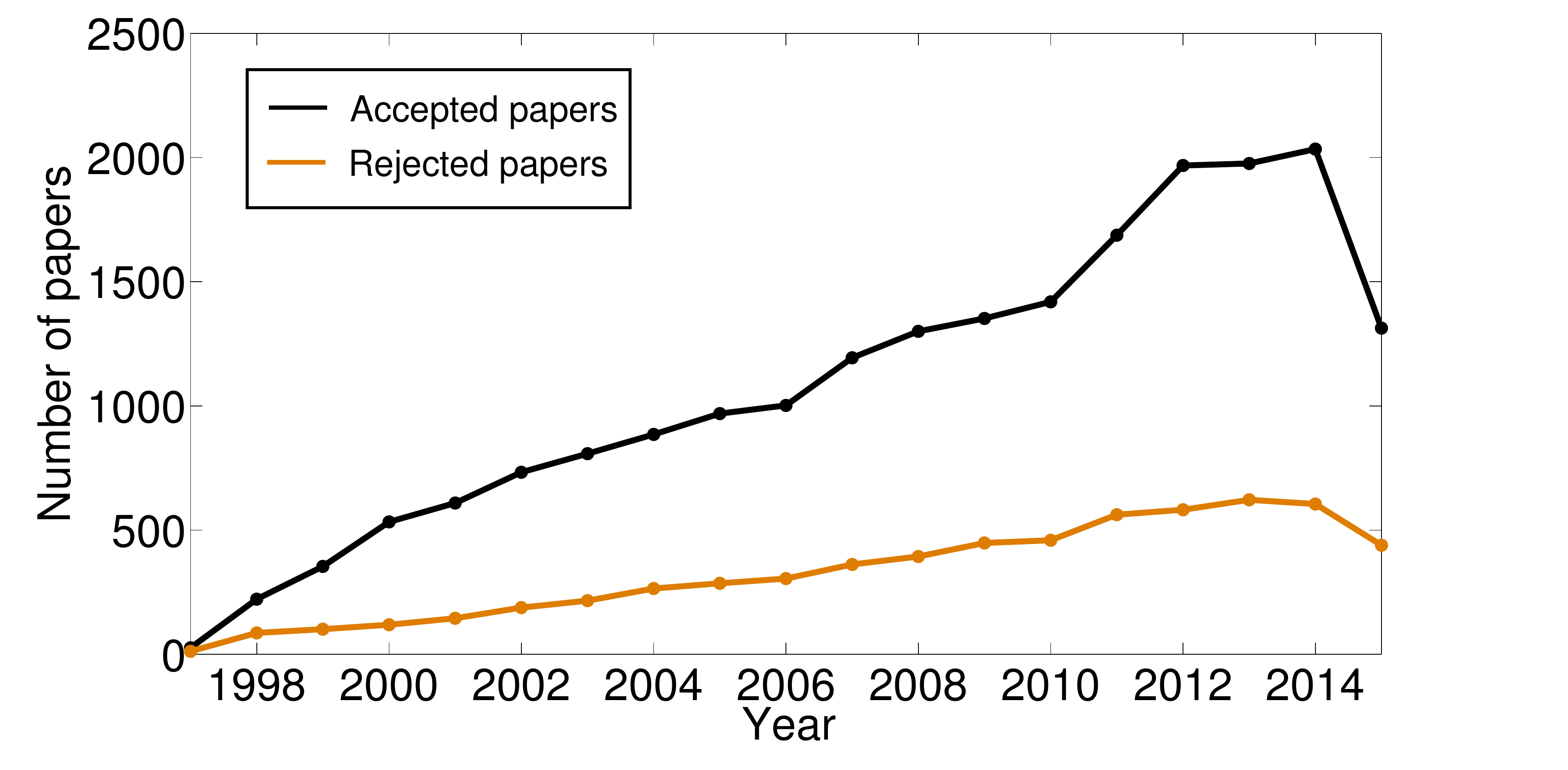} & \includegraphics[scale=.15]{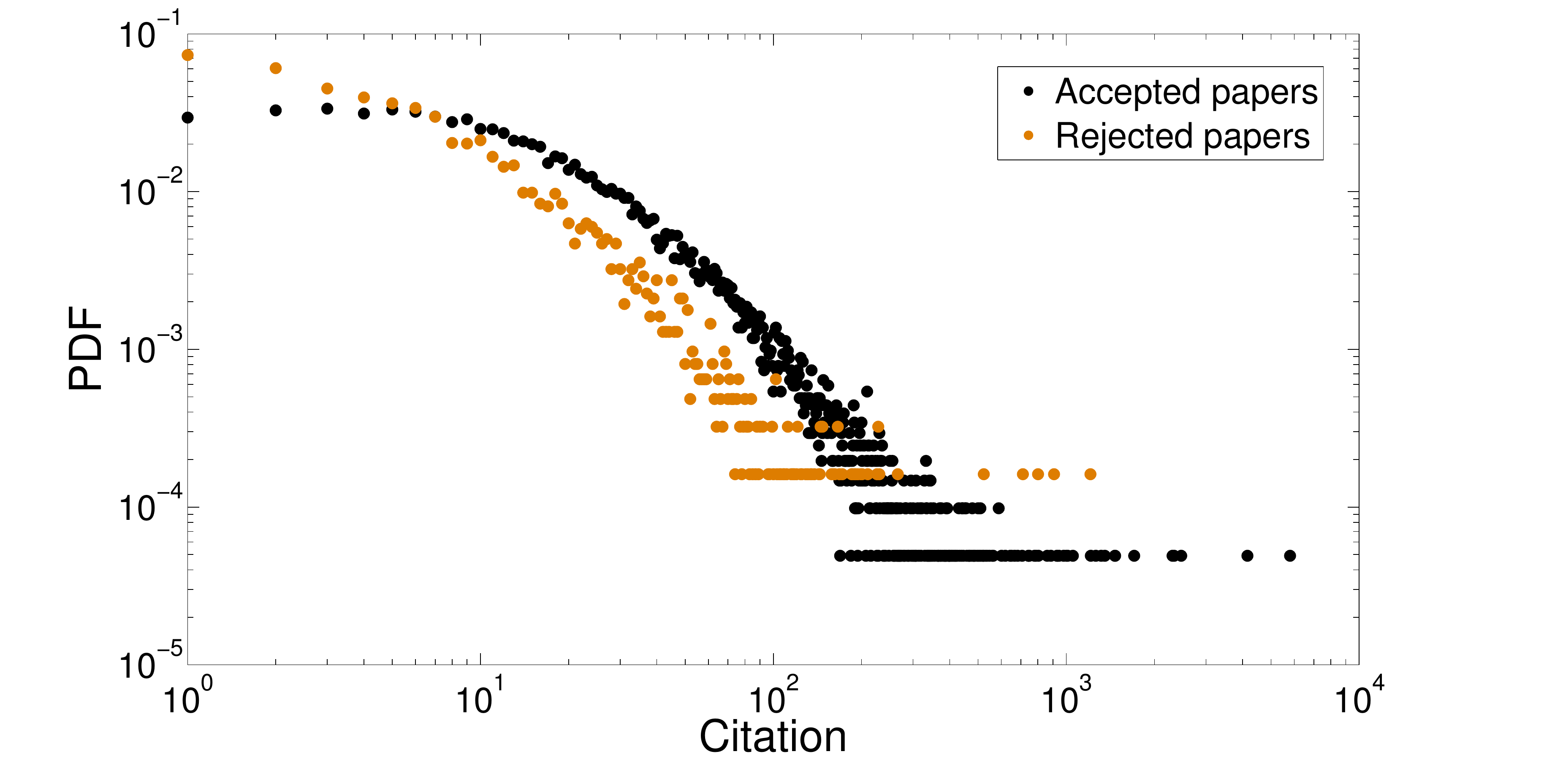} & \includegraphics[scale=0.15]{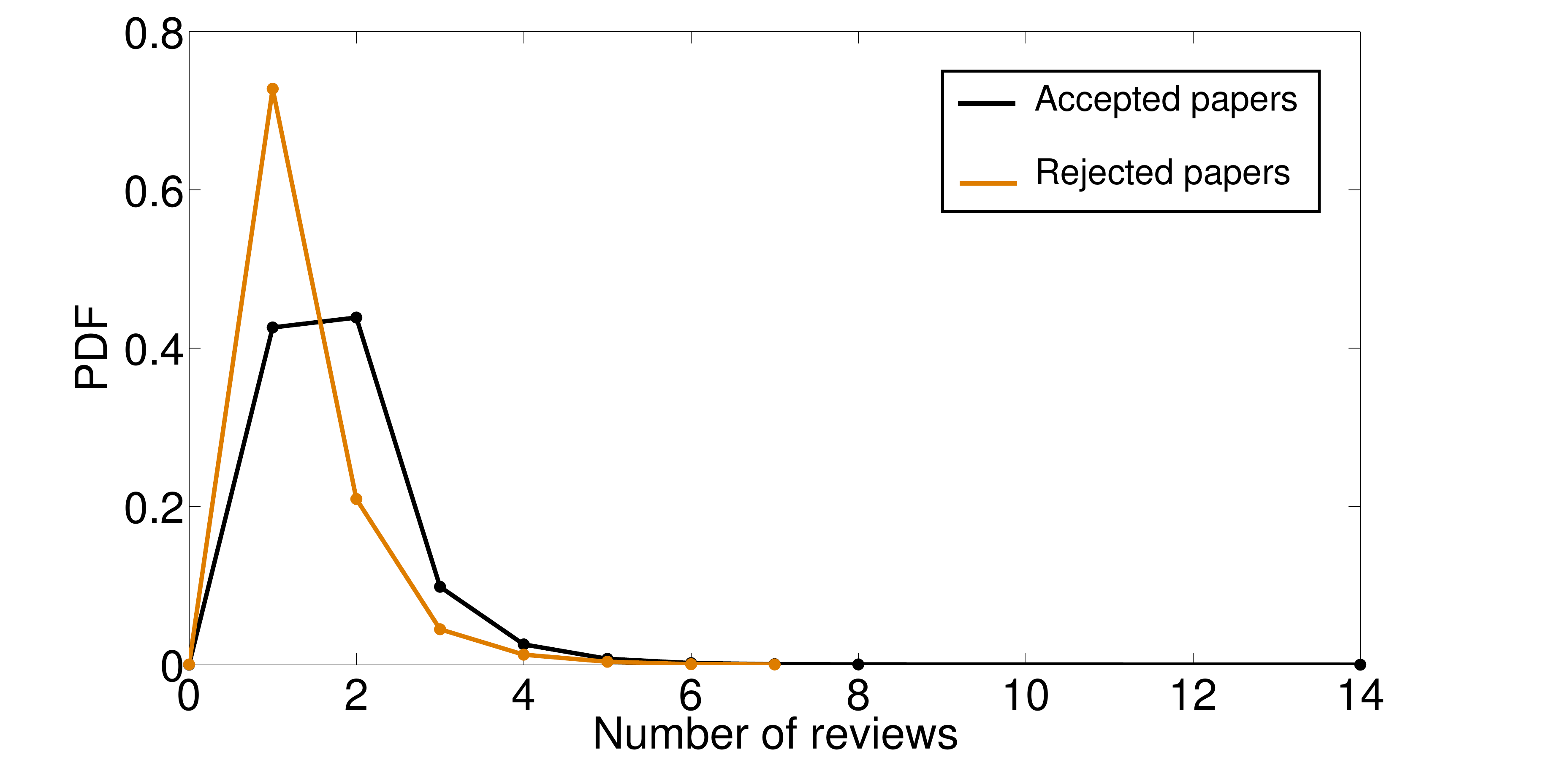}
\end{tabular}
\caption{{\bf (Left)} Number of accepted and rejected papers per year from 1997 to 2015. {\bf (Middle)} Citation distribution of both the accepted and the rejected papers. {\bf (Right)} Distribution of number of reviews for accepted and rejected papers.}
\label{fig1}
\end{figure*}

\noindent{\bf Determining the fate of the paper:} 
Based on the network features built above, we propose a supervised model which quite accurately predicts ($R^2$ = {\bf 0.79}, $RMSE$ = {\bf 0.496}) the long-term citation of a paper. In addition, if we also include the supporting features into the model we obtain further gains ($R^2$ = {\bf 0.81}, $RMSE$ = {\bf 0.46}). Analysis of the importance of the features shows that the network features are the strongest predictors for this task.
We believe that our system would be of immense help in assisting the editor in deciding acceptance or rejection of the paper specifically in cases when the review reports are contradictory. 
Note that while our work is a case study of JHEP, the formulations that we report are very general and can be extended to any other available dataset. 

\section{Related works}
\label{related_works}

The importance and effectiveness of peer-review have long been studied across diverse platforms. The benefits of collaborative learning has been studied extensively in~\cite{reily2009two}. In fact the authors find that peer reviews are accurate compared to accepted evaluation standard. The benefits of peer-review in classroom environment has been summarized in~\cite{hamer2005method}. The social media has also been using peer-review to filter information~\cite{diakopoulos2011towards,lampe2004slash}.  
In fact Wikipedia, Reddit and Quara rely largely on peer-review for improving the quality of the content \cite{stvilia2008information}. Peer-review in online collaboration context has also been studied in great detail \cite{brindley2009creating, huang2016effectiveness}.  

In the scientific community also peer-review system has long been relied upon for preventing flawed research from getting into the literature ~\cite{kassirer1994peer}. However, several concerns about the limitations of this system have been conveyed~\cite{ingelfinger1974peer,relman1989good,smith2006peer} in the context of scientific peer-review. The most significant of them is the presence of bias while evaluating the quality of the research~\cite{lee2013bias}.  In~\cite{mcnutt1990effects}, authors showed that blinding significantly improved the quality of the reviews. Further, reviewers seem to fail to reach consensus while determining whether the contribution of a scientific article is novel enough~\cite{cole1981chance}. 

This has led to serious debate in the research community regarding the credibility of the peer-review process. With the total number of published articles reaching $1.3$ million in 2006 (~\cite{bjork2009scientific}) and expected to grow at even faster rate, we believe a thorough analysis of the reliability of the peer-review process is needed since a large amount of time as well as cost is involved in this process.

The relationship between reviews and citations for a paper has not been studied previously to the best of our knowledge except for some preliminary attempts in~\cite{braatz2014papers, coupe2013peer}.
In~\cite{coupe2013peer} it has been showed that the papers that are declared as best papers at a publication venue end up getting lesser number of citations while~\cite{braatz2014papers} points out that rejected papers tend to get more citations in the long run.

Although there are several works on determining usefulness of user supplied reviews especially in the e-commerce domain~\cite{ghose2011estimating, kim2006automatically}, similar analysis for review reports of scientific articles have remained mostly unexplored. We believe that our findings will not only re-establish the need for a well-moderated peer-review process but also provide important insights as to how such a supportive process could be further improved. In this paper, we consider a very rich dataset to address some of the issues that remained unattended so far in the literature. The observations that we make are very unique and the conclusions that we draw, thereby, are significantly novel adding huge value to the rich digital library literature.

\section{Dataset}
\label{dataset}

The main aim of this work is to understand the importance of the review process and especially the contribution of the referees in this process. Such an analysis requires detailed information of reviews like the number of rounds of review, final decisions and the text of the review report in each round and, lastly, the number of citations for each paper. 

{\em Obtained data:} As stated earlier, for our analysis, we consider the dataset of the {\em Journal of High Energy Physics} (JHEP). It is one of the leading journals in its field and publishes theoretical, experimental and phenomenological papers. Among JHEP's most direct competitors there are Physical Review D, Nuclear Physics, EPJC, Physics Letters B and Physical Review Letters.
This dataset consists of {\bf 28871} papers that were submitted between 1997 (year of inception) and 2015 of which {\bf 20384} were accepted and {\bf 7073} were rejected. The rest of the papers were either withdrawn by the authors or the final decisions were not available. The number of distinct review reports is {\bf 70k} containing {\bf 12m} lines of review text. For each paper we have the title, the abstract, the authors, the date of publication (in case it was accepted) and the number of citations for the accepted papers. The dataset further contains for each paper the number of rounds of reviews it received before it was accepted (rejected) as well as the detailed text of the report submitted by the assigned reviewer and the editor.

\noindent{\em Pre-processing:} To obtain the necessary information for the rejected papers we queried the ``Inspire'' search engine\footnote{\url{https://inspirehep.net}} by their corresponding arXiv\footnote{\url{http://arxiv.org/}} ids. We could obtain for each paper the citation information, the abstract, the title, the authors and also the publishing journal (if at all it got published). Note that all through our analysis we refer to number of citations as the cumulative number of citations that a paper/author obtained at the end of 2015. 
We further had to disambiguate the names of the authors and assign each of them a unique id.  

\noindent{\em Some basic facts about the dataset:} In fig.~\ref{fig1}{\bf(Left)} we plot the year-wise distribution of the accepted and the rejected papers from 1997 to 2015. We observe an increasing trend in the number of submissions except for the year 2015 for which the data is incomplete. 
\begin{table}[htpb]
\centering
\caption{General information of the dataset.}
\label{tab1}
\resizebox{!}{1.4cm}{
\begin{tabular}{|l|l|}
\hline
Number of papers                                                                 & 28871 \\ \hline
\hline
\begin{tabular}[c]{@{}l@{}}Number of papers (accepted)\end{tabular}            & 20384 \\ \hline
\begin{tabular}[c]{@{}l@{}}Number of papers (rejected)\end{tabular}            & 7073  \\ \hline
\begin{tabular}[c]{@{}l@{}}Average number of reviews (accepted papers)\end{tabular}   & 1.76  \\ \hline
\begin{tabular}[c]{@{}l@{}}Average number of reviews (rejected papers)\end{tabular}   & 1.35  \\ \hline
\begin{tabular}[c]{@{}l@{}}Average number of citations (accepted papers)\end{tabular} & 31.89 \\ \hline
\begin{tabular}[c]{@{}l@{}}Average number of citations (rejected papers)\end{tabular} & 9.45  \\ \hline
\end{tabular}}
\end{table}

In fig.~\ref{fig1}{\bf(Middle)} we plot the citation distribution of the accepted and the rejected papers. Both the distributions seem to follow a power-law behavior. We further plot 
the distribution of the number of reviews for the accepted and the rejected papers in fig.~\ref{fig1}{\bf(Right)}. An important observation is that for JHEP, majority of the 
papers undergo one or two rounds of reviews after which they are either accepted or rejected. 
We note certain general information related to the dataset in table~\ref{tab1}.
Each submitted paper in the dataset also consists of the list of authors. There are {\bf 15127} unique authors in the dataset with at least one submission to JHEP and {\bf 12434} authors with at least one accepted paper. The average number of submissions per author is {\bf 5.18} while the number of authors per paper is {\bf 2.87}. 

\begin{figure}
\centering
\includegraphics[scale=0.3]{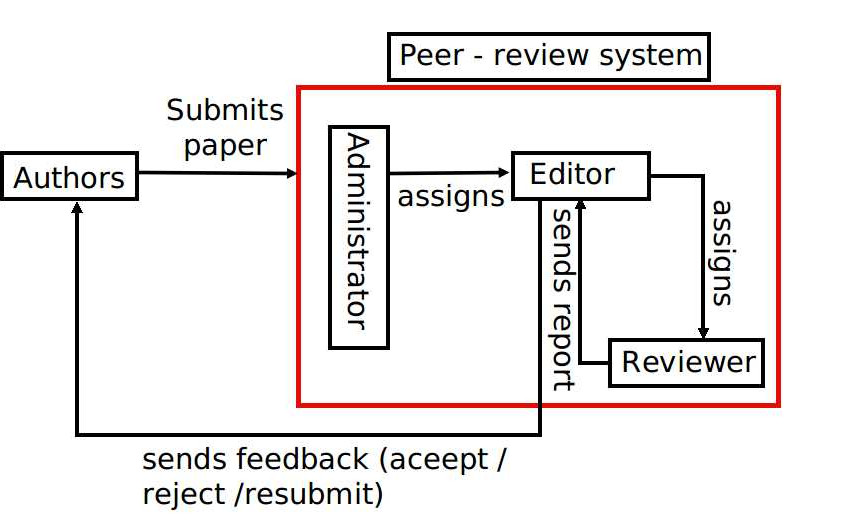}
\caption{\label{peer_review} Peer-review system in JHEP}
\end{figure}

\noindent{\em Peer-review process in JHEP:} For every submitted paper the administrator assigns an editor for it. The editor selects a single or a small set of reviewers for judging the quality of the contributions in the paper. The reviewer sends back his views in the form of a report. Based on this report the editor decides whether to accept or reject the paper. The editor may also ask the authors to reshape the paper based on the feedback of the reviewer(s) and in which case they have to resubmit before a decision on its acceptance could be taken. In fig.~\ref{peer_review}, we present a schematic showing the peer-review process in JHEP.

\section{Reviewer-reviewer interaction network}
\label{rev_int_net}
In this section, we construct a reviewer-reviewer interaction network and show that its properties are linked to the future scientific impact of a paper (measured in terms of the cumulative citation count). In specific, we find that the position of the assigned reviewer in the network (measured in terms of degree, centrality, clustering coefficient and PageRank) could be used to predict the long term citation of the paper. 
 
The reviewer-reviewer interaction network is created with each node representing a reviewer and an edge exists between two reviewers if they have been assigned by at least one common editor. We devote the rest of the section in demonstrating the importance of the various structural properties of the network in determining the long term citation of the paper. Note that there are 4035 unique reviewers in the system each of which form a node in this network.

\begin{figure*}
\centering
\includegraphics[width = 0.95\textwidth]{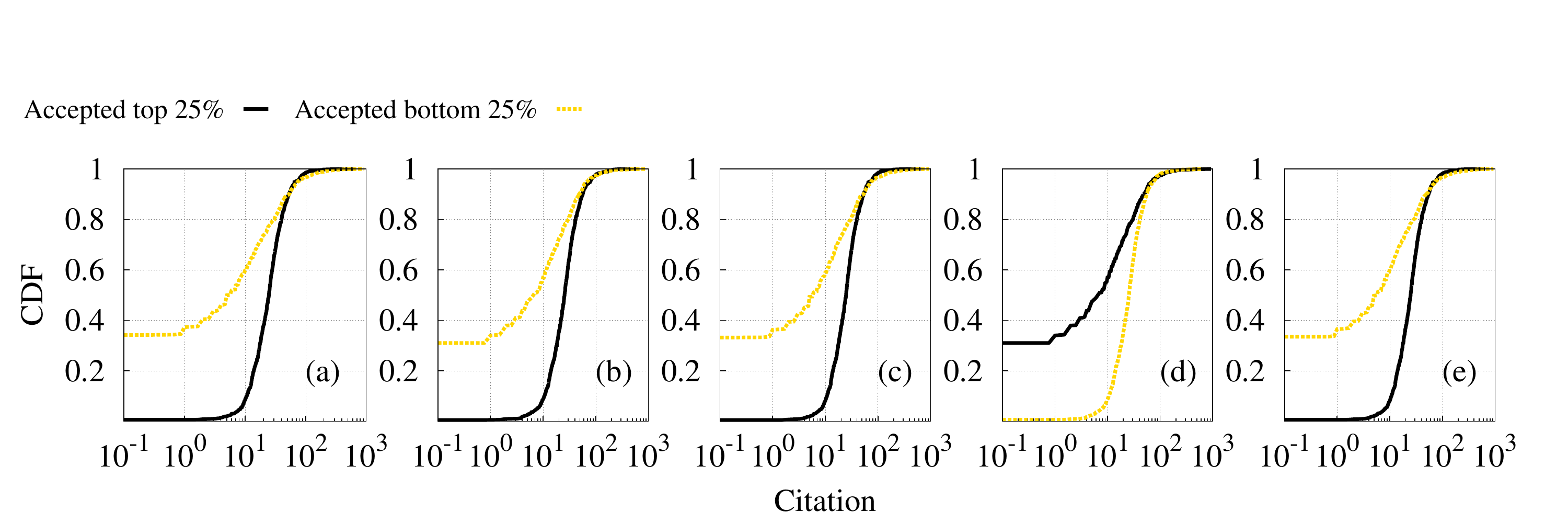} 
\caption{\label{fig:net_citation} Cumulative distribution function (CDF) of citations received by the papers (accepted) reviewed by referees in top 25\% and bottom 25\% reviewers ranked according to (a) degree, (b) betweenness centrality, (c) closeness centrality (d) clustering coefficient values and (e) PageRank in the reviewer-reviewer interaction network.}
\end{figure*}

\subsection{Degree (Deg)}
Degree of a node $v$ is the number of other nodes it is connected to in the network. A node with a higher degree in the reviewer-reviewer interaction network would indicate (i) assignment from multiple editors, (ii) assignment from a reputed editor (with large number of assignments) which in turn would indicate the reputation of the reviewer. To verify our hypothesis, we rank the reviewers based on their degree in the network and calculate the mean citation of the papers reviewed by the reviewers in the top and the bottom 25\% of the rank list. We observe that  the papers reviewed by the top 25\% reviewers receive much higher citations than the those reviewed by the bottom 25\% reviewers (refer to fig.~\ref{fig:net_citation}(a)).    

\subsection{Betweenness centrality (BC)}

Betweenness centrality of a node quantifies the position of a node based on the number of shortest paths the node is part of. For every  pair of nodes in the network there exists a shortest path between them. Betweenness centrality of a node ($v$) is the fraction of all such paths that pass through $v$. In the reviewer-reviewer interaction network, a high  centrality value would indicate assignment by multiple editors and that this node acts as a bridge between them. We again rank the reviewers based on the betweenness centrality values and calculate the average citation of the papers. We find that the papers accepted by the top 25\% reviewers tend to be cited more compared to those accepted by the bottom 25\% (refer to fig.~\ref{fig:net_citation}(b)). 

\subsection{Closeness centrality (CC)}

Formally closeness centrality of a node in a network is the inverse of the sum of length of its shortest path to all other nodes in the network. Hence higher centrality value indicates that the node is more closer to all other nodes in the network. In the reviewer-reviewer interaction network, a reputed reviewer will be assigned by multiple reviewers and hence will be closer to the other reviewers in the network. This is represented in fig.~\ref{fig:net_citation}(c), where we show that the papers accepted by top 25\% most central reviewers 
are cited more often compared to the bottom 25\% reviewers. 

\subsection{Clustering coefficient (Clus)}

Clustering coefficient of a node is measured as the fraction of connections among the neighbors of the node. For the reviewer-reviewer interaction network, every reviewer assigned by a common editor is connected to every other reviewer in the network. A reviewer assigned by many editors would actually act as a bridge between two cliques and hence would have a lower clustering coefficient value compared to a reviewer who is part of a single clique (always assigned by a single editor). This is further demonstrated in  fig.~\ref{fig:net_citation}(d) where we observe that the papers accepted by reviewers having lower clustering coefficient tend to be cited more.

\subsection{PageRank (PR)}

PageRank is a link analysis based algorithm that calculates for each node its relative importance within 
the network. Specifically, PageRank outputs a probability distribution which is used as the likelihood of a random walker to end up in a specific node. We simulate PageRank on the reviewer-reviewer interaction 
network to obtain the relative importance of each node. Further analysis indicates that the papers accepted by the top 25\% reviewers (based on PageRank) are cited more often compared to those accepted by the bottom 25\% reviewers (refer to fig.~\ref{fig:net_citation}(e)). 

The above results thus indicate that simple network properties of the reviewer-reviewer interaction network could be highly effective in predicting the long-term citation of the paper at the time of publishing. 



\section{Supporting features}
Most of the works~\cite{yan2012better,chakraborty2014towards} related to predicting long-term citation of papers considers a wide set of author related features, papers-centric features and citation pattern of the paper in the first few years from the date of its publication. 
Further, our dataset allows us (unlike the existing datasets) to look into several other features related to the review-process like the review report, the behavior of the assigned referee, the number of rounds of reviews the paper went through and others. 
We consider the above features as well as those existing in the previous literature (wherever available) as the set of supporting features. The features are categorized into (i) paper based, (ii) review report based, (iii) author based and (iv) reviewer based. Apart from investigating the effectiveness of a feature in determining the long-term citation of the paper, we also point out some interesting observations that we could make while analyzing the dataset.    

\subsection{Paper based features}
\label{analysis}
We have already observed that the average number of citations received by accepted papers is 31.80; for rejected papers the corresponding value is 9.45 (refer to table~\ref{tab1}). Note that for each paper we consider the citations accrued by it till 2015 from the date of publication.
We further consider all the accepted and the rejected papers and segregate them based on the number of citations received. We consider different citation buckets and plot the fraction of accepted and rejected papers in each of these buckets in fig.~\ref{fig4}{\bf (Left)}. Note that the bucket sizes are in increasing powers of 2. Typically, the buckets are $\leq 1$, $2$, $(>2$ and $\leq 4)$ and so on. It can be clearly observed that accepted papers are cited more often compared to the rejected ones. Nevertheless, on further analysis we find that there could be a few exception cases where the rejected paper could make a place in higher impact journals (compared to JHEP) and, eventually, receive a high volume of citations in future. We present two such pathological cases below -- {\bf Case 1:} Rejected after two rounds of review, later accepted at Physics Letters B, citations: 1209; {\bf Case 2:} Rejected after one round of review, later accepted at Computer Physics Communications, citations: 929. We perform a thorough analysis of these irregular cases later in the paper.

\subsubsection{Number of review rounds (RR)}

 \begin{figure*}[htpb]
 \centering
 \begin{tabular}{ccc}
 \includegraphics[scale=0.15]{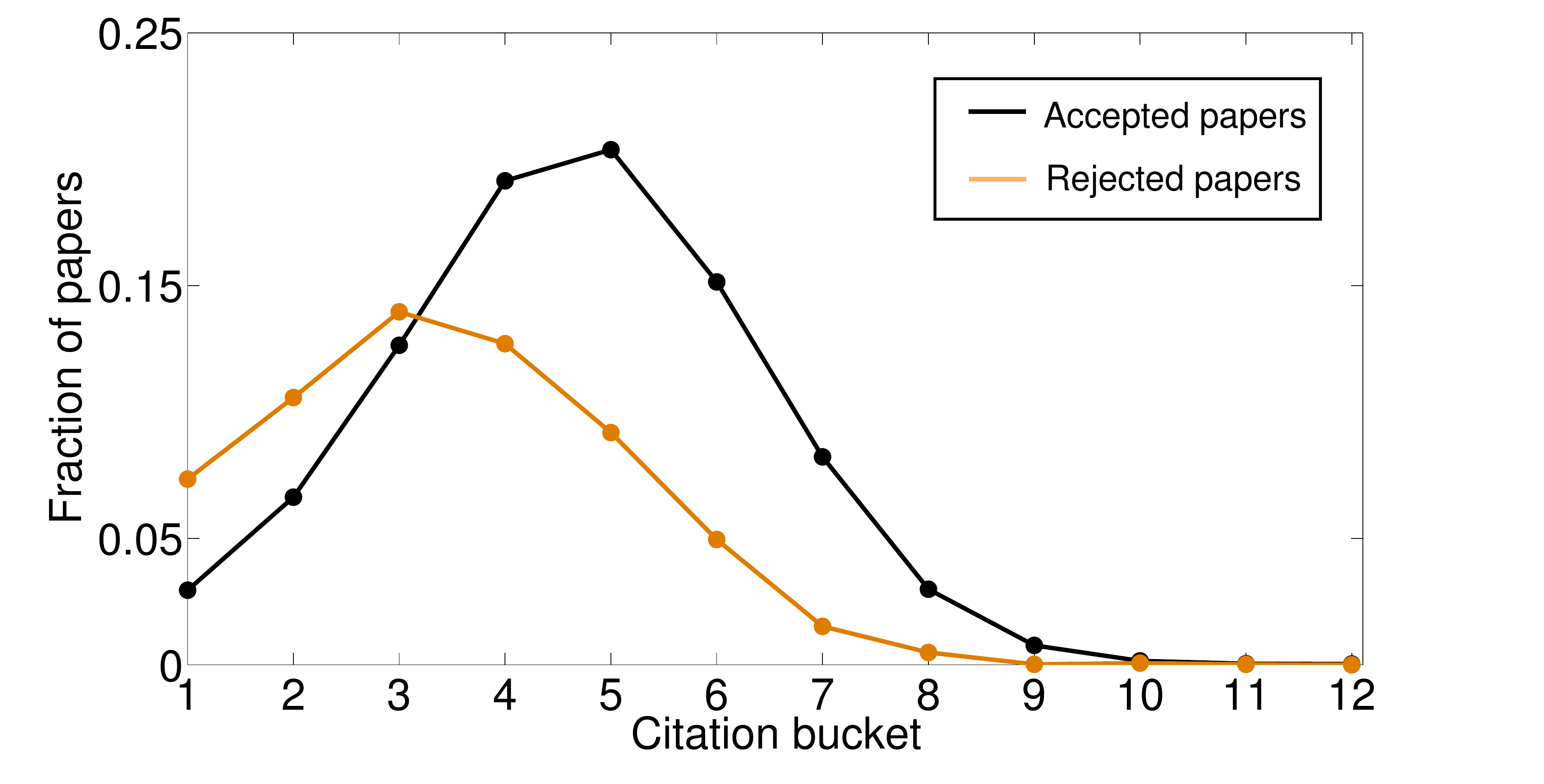} & \includegraphics[scale=0.15]{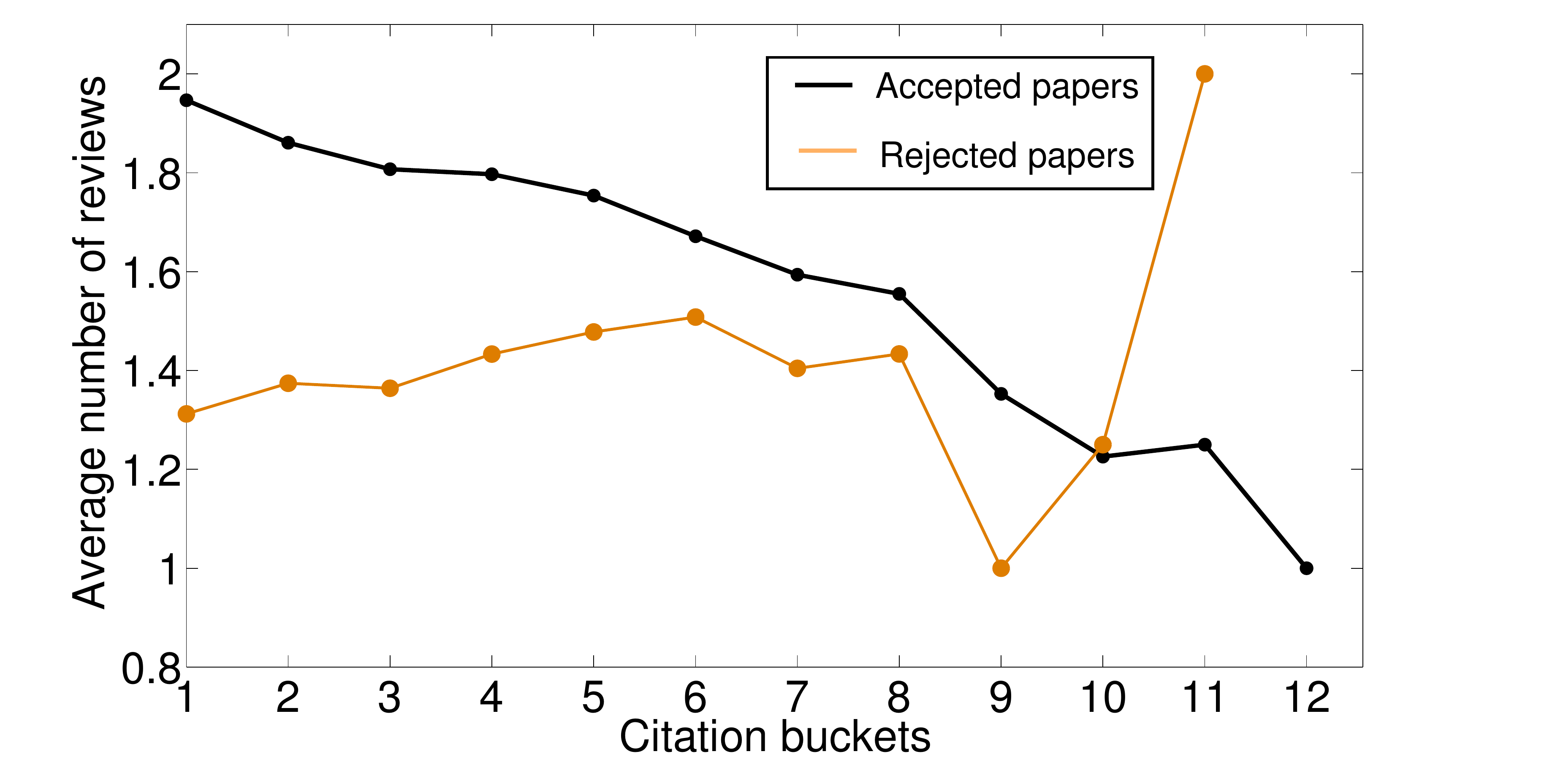} & \includegraphics[scale=0.15]{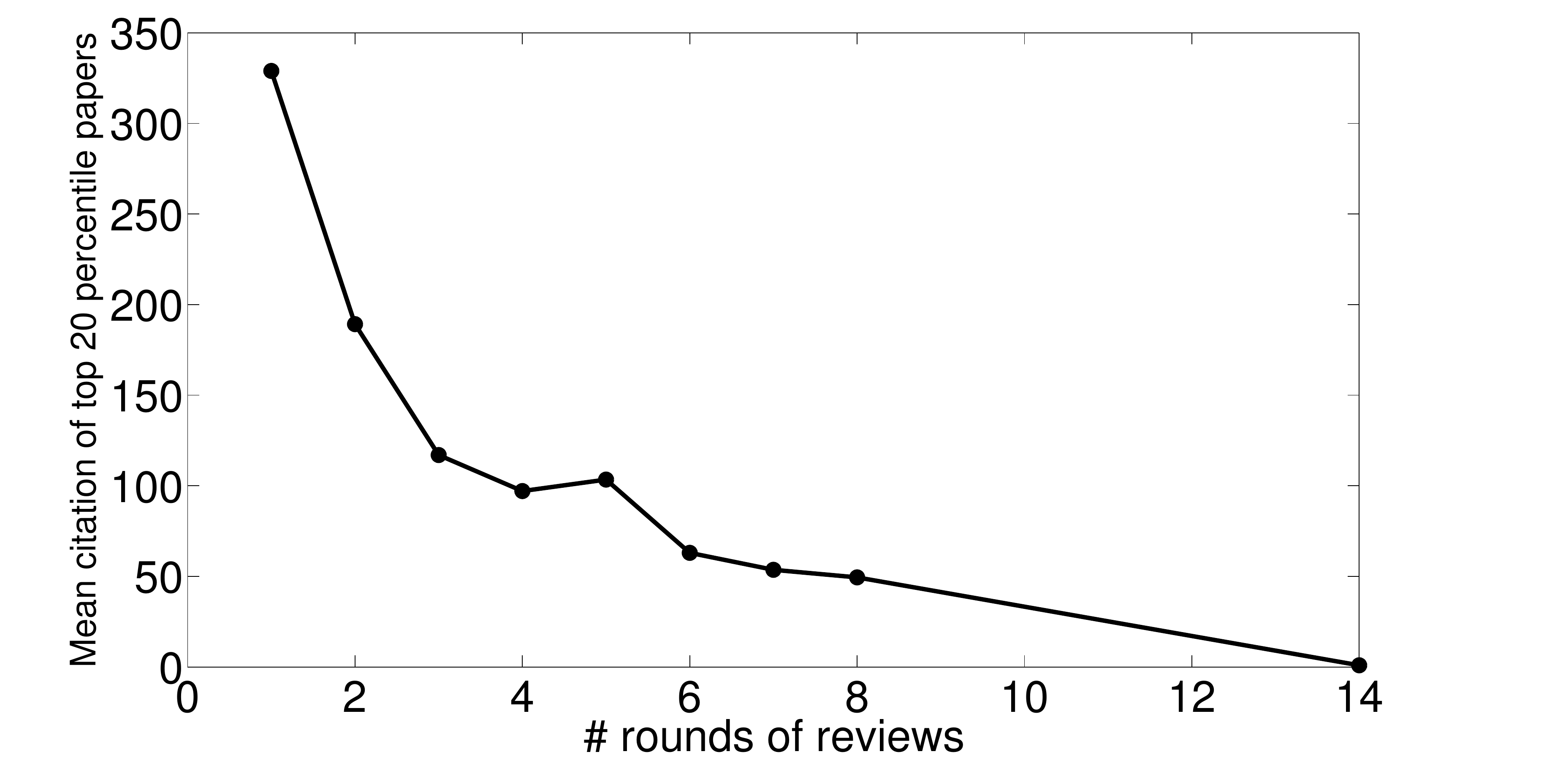}
 \end{tabular}
  \caption{{\bf (Left)} Fraction of accepted and rejected papers in different citation buckets. {\bf (Middle)} Average number of reviews for accepted and rejected papers in different citation buckets. For both {\bf (Left)} and {\bf(Middle) } bucket sizes are in increasing powers of 2. E.g. $\leq 1$, $2$, ($>2$ and $\leq 4$) and so on. {\bf (Right)} Average citation of the top 20 percentile  papers for a given number of rounds of review request.}
   \label{fig4}
 \end{figure*}

We next check whether the {\bf review rounds} improve the quality of the paper. To this aim we segregate the papers based on the number of citations into different buckets and for each bucket we calculate the average number of reviews the papers received in that bucket. The bucket sizes are again in increasing powers of 2. Typically, the buckets are $\leq 1$, $2$, $(>2$ and $\leq 4)$ and so on. We plot the results in fig.~\ref{fig4}{\bf (Middle)}. We observe that for accepted papers, the low cited ones on average tend to get accepted after more rounds of reviews while the high cited ones undergo lesser rounds of reviews before getting accepted. It can hence be concluded that going through the review process multiple times does not necessarily improve the quality of the papers much that are eventually accepted. For the rejected papers we observe a contrasting trend indicating that the review process indeed helped in improving the quality of the paper in the long run possibly enhancing the chances of its acceptance at a different venue later. For the accepted papers we further classify them based on the number of reviews they received and calculate the average citations of the top 20 percentile papers (ranked by citations) in each class (refer to figure~\ref{fig4}{\bf (Right)}). We observe that the average citation drops as the number of reviews increases further suggesting that papers accepted after higher number of reviews often fail to create a high citation impact. This indicates that the number of rounds of review could be an indicator of the long-term citation of the paper. 

\begin{figure}
\centering
\includegraphics[scale  = 0.25]{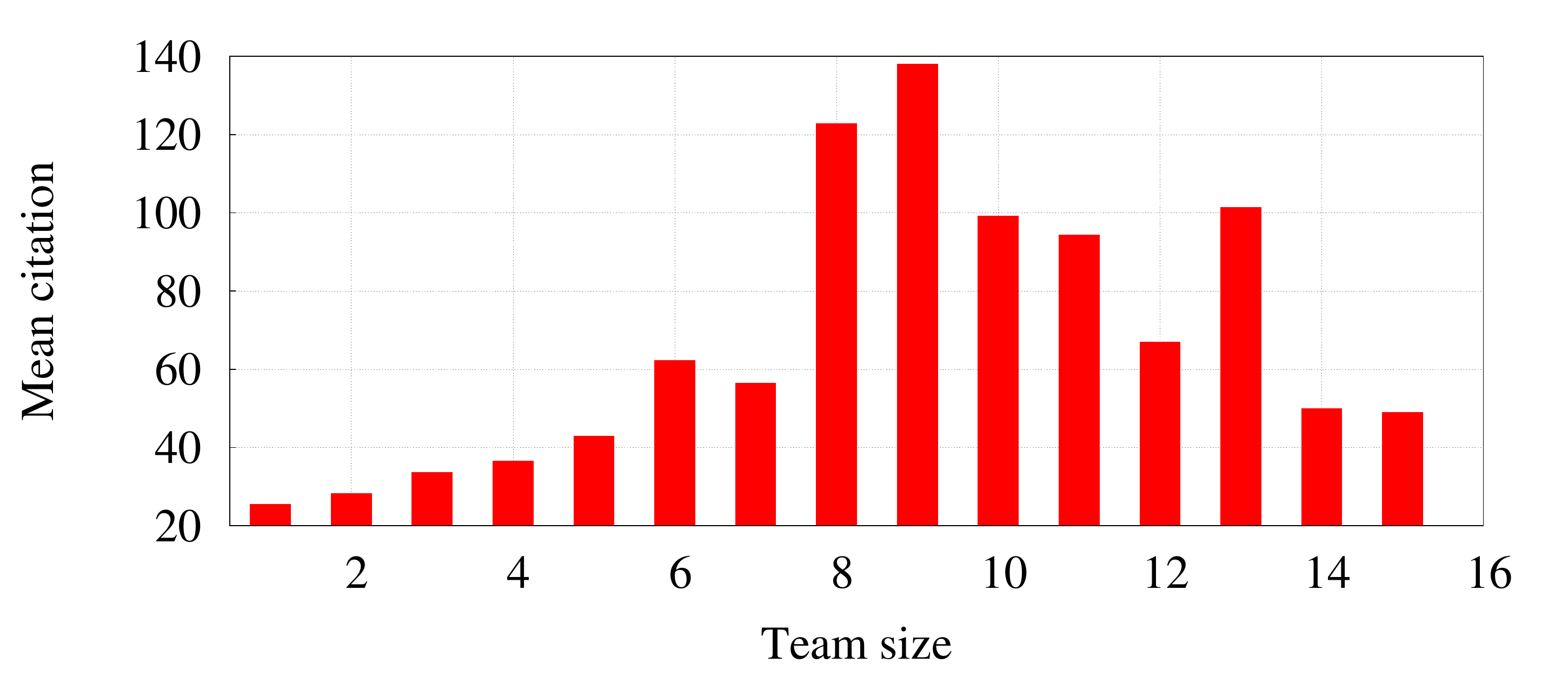}
\caption{\label{team:citation} Average citation versus team size. Note that we segregate the papers based on the teams size and calculate the average citation.}
\end{figure}

\subsubsection{Team size (TS)}
The authors in~\cite{chakraborty2014towards} hypothesized that there exits an optimal team size for which the citations received by the paper is maximum. We hence segregate the papers based on the team size and calculate the mean citation of the papers. We observe that team size 9 (refer to fig.~\ref{team:citation}) is the optimal as the papers with 9 authors gets more citation on average.


\begin{figure}
\centering
\includegraphics[scale=0.25]{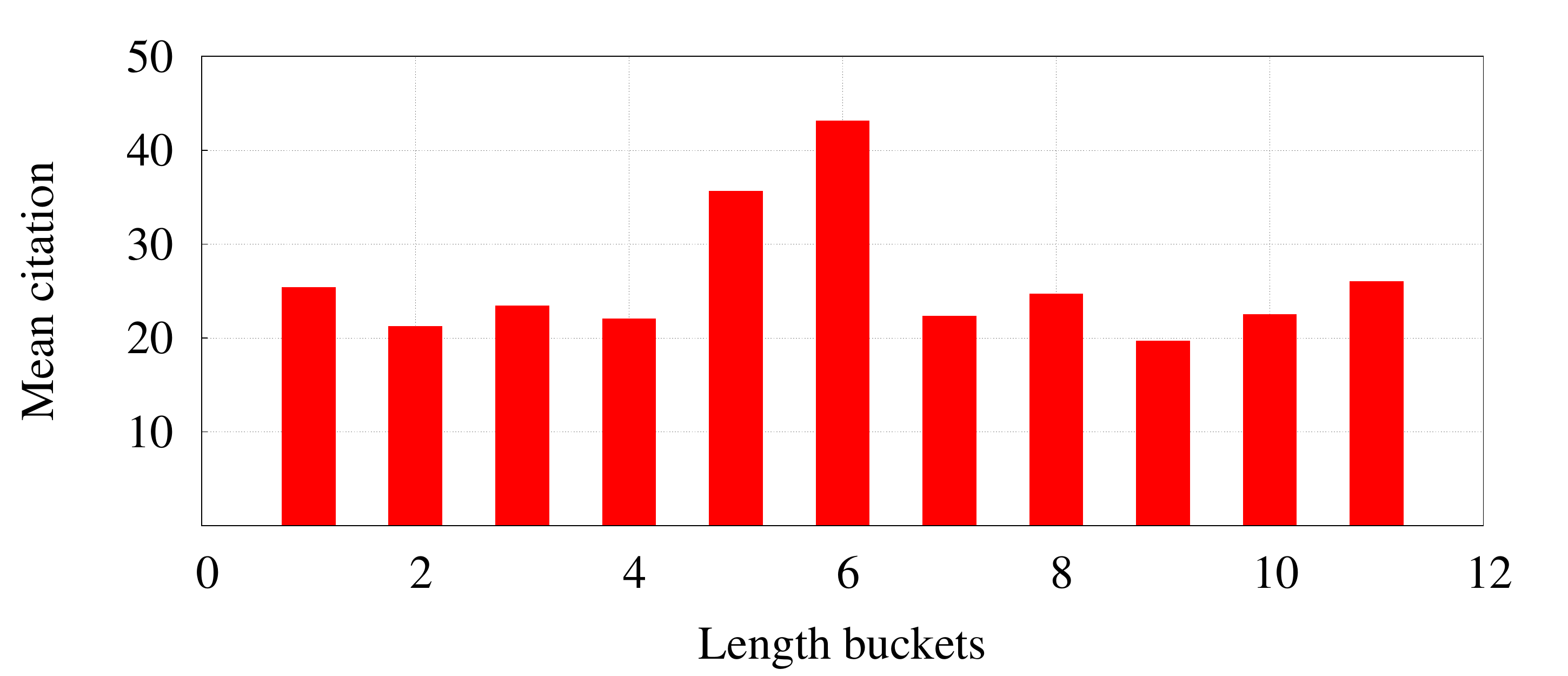}
\caption{Mean length of referee reports in terms of number of words at different rounds of review. Typically the buckets are $< 100$, ($\geq 100, < 200$) and so on}
\label{fig:length}
\end{figure}

\subsection{Review report based features}
\label{text_analysis}

In this subsection we analyze whether certain features could be extracted from the reports sent by the reviewers that could be an indicator of the long-term citation of the paper. Note that we have two types of reports -- {\bf Referee report}: Report sent by the assigned referee to the editors and {\bf Editor report}: Report sent by the editor to the authors based on the referee report. We primarily focus on the referee reports as editorial reports are in almost all cases a reiteration of the referee reports.

\subsubsection{Length of the reports (RL)}
We start by looking whether the length of the review reports sent by the reviewers are indicative of the quality and hence the long-term citation of the paper. To this aim we segregate the papers based on the length of the report and calculate the mean citation of each of these buckets. The lengths are bucketed with sizes typically $< 100$, ($\geq 100, < 200$) and so on. We observe that there exists an optimal length (between 500 and 600 words) for which the citation obtained by the corresponding paper is maximum  (refer to figure \ref{fig:length}).

\subsubsection{Sentiments (SNT)}
We next perform sentiment analysis on the review reports. To determine the sentiment of a report we use a method described in~\cite{montejo2012random} which performs a graph-based word sense disambiguation and lexical similarity analysis using a pre-existing knowledge base. A sentiment score of 0 indicates that the document is neutral, a positive score  indicates a positive sentiment and a negative score indicates a negative sentiment.

\begin{figure}[htpb]
\centering
\includegraphics[scale=0.23]{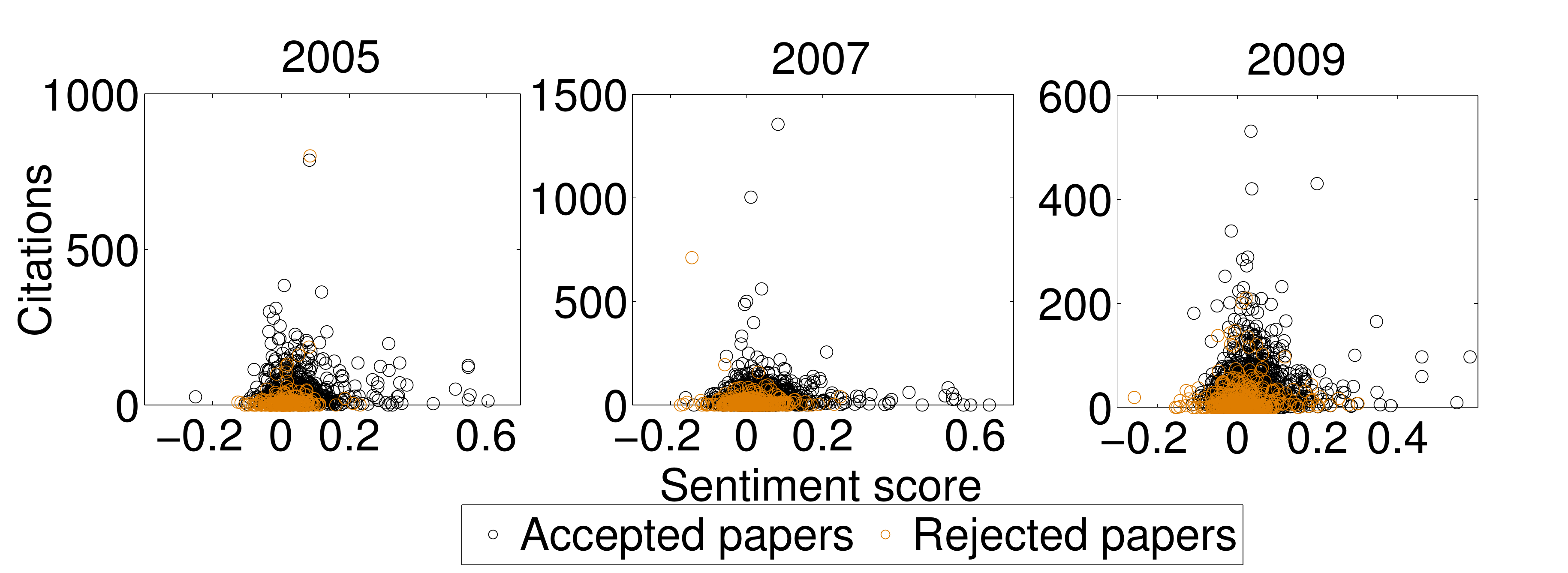}
\caption{Sentiment score versus citations for both accepted and rejected papers for the years 2005, 2007 and 2009. We find similar trends for other years as well.}
\label{fig12}
\end{figure}

To determine whether the overall sentiment of the reviews of a paper is related to the number of citations received by it, we plot for each paper (both accepted and rejected) the number of citations it received against the  sentiment score of the first round of review in fig.~\ref{fig12}. Note that we segregate the papers based on the year of the publication or the rejection. We observe that the highly cited papers mostly have reviews with neutral or positive sentiment. Accepted papers with positive reviews are, on average, found to receive $25.27$ citations while those with negative reviews are found to receive $13.56$ citations.
However, there are cases where the accepted paper received highly positive reviews but was not cited. Conversely, there are cases where the sentiment was neutral but the paper garnered a large number of citations. Shown below are two such examples -- {\bf Case 1:} Year of publication: 2006, sentiment score: 0.0234 (almost neutral), citations: 5812; {\bf Case 2:} Year of publication: 2008, sentiment score: 0.65 (highly positive), citations: 6.

For the rejected papers we observe that those which received neutral reviews but were rejected, tend to garner higher citations later compared to the ones which received negative reviews. There are certain exceptions as well, two of which are -- {\bf Case 1:} Year of rejection: 2010, sentiment score: 0.27 (positive), citations: 1; {\bf Case 2:} Year of rejection: 2007, sentiment score: -0.14 (negative), citations: 711.
Manual investigation of the review text shows that the papers which are highly cited after rejection were mainly rejected for not being in the scope of JHEP and not because of flawed results. 

\subsubsection{Linguistic quality indicators (LQI)}

Here we check whether there are linguistic quality features present in the review reports which can serve as an indicator of the future impact of the paper. To our aim we use the LIWC\footnote{\url{http://liwc.wpengine.com/}} (Linguistic Inquiry and Word Count) text analysis tool~\cite{pennebaker2007development}. The tool provides, as output, percentage of words in different categories for an input text. The categories are broadly divided into linguistic (21 dimensions like pronouns, articles etc.), psychological (41 dimensions like affect, cognition etc.), personal concern (6 dimensions), informal language markers and punctuation apart from some general features like word count, words per sentence etc. We apply the LIWC tool on the review reports for our analysis and mainly focus on the linguistic and the psychological categories.
Next we check whether the LIWC features discussed earlier can also serve as indicators differentiating high and low cited papers. We rank the papers based on the number of citations they have received and consider the top 10\% as highly cited and the bottom 10\% as low cited papers. Note that we only consider the papers that were published before 2012 so that the papers have at least three years of citation history. In table~\ref{tab3} we report the mean percentage of words in different LIWC categories across all the papers (both high and low cited). We find several quality indicators here as well. The key observations are: (i) future tense is used more significantly in case of review reports of highly cited papers compared to low cited papers. On manually investigating the reviews of some of the highly cited papers we observe that statements like ``its result will become a useful addition to ..'' are prevalent; (ii) insightful and inclusive words are also used to a greater extent in review reports of highly cited papers compared to low cited papers; (iii) positive words are also more prevalent in highly cited papers as well.\\
Thus, these indicators show that the reviewers were, in many cases, indeed able to guess the quality of the paper as is evident from the review reports.

\begin{table}
\centering
\caption{Mean values of percentages of various categories of words in review reports of high and low cited papers where the means differ significantly.}
\label{tab3}
\resizebox{4.5cm}{!}{
\begin{tabular}{|l|l|l|l|}
\hline
Category                    & Dimension                                                   & \begin{tabular}[c]{@{}l@{}}High cited\\ papers\end{tabular} & \begin{tabular}[c]{@{}l@{}}Low cited\\ papers\end{tabular} \\ \hline \hline
\multirow{2}{*}{Linguistic} & Future tense                                                & 1.17                                                          & 1.05                                                         \\ \cline{2-4} 
                            & Negation                                                    & 0.72                                                          & 0.84                                                         \\ \hline
\multirow{4}{*}{Cognitive}  & Insight                                                     & 3.52                                                          & 3.16                                                         \\ \cline{2-4} 
                            & Causation                                                   & 2.60                                                          & 2.38                                                         \\ \cline{2-4} 
                            & Inclusive                                                   & 3.70                                                          & 3.43                                                         \\ \cline{2-4} 
                            & Exclusive                                                   & 1.28                                                          & 1.52                                                         \\ \hline
Affective                   & \begin{tabular}[c]{@{}l@{}}Positive \\ emotion\end{tabular} & 2.84                                                          & 2.70                                                         \\ \hline
\end{tabular}}
\end{table}

\subsection{Author based features}
\label{author_analysis}

We next look into some of the author based features like author reputation and author productivity to determine how they influence the long-term citation of the paper.  
\begin{figure*}
\centering
\begin{tabular}{ccc}
\includegraphics[scale=0.15]{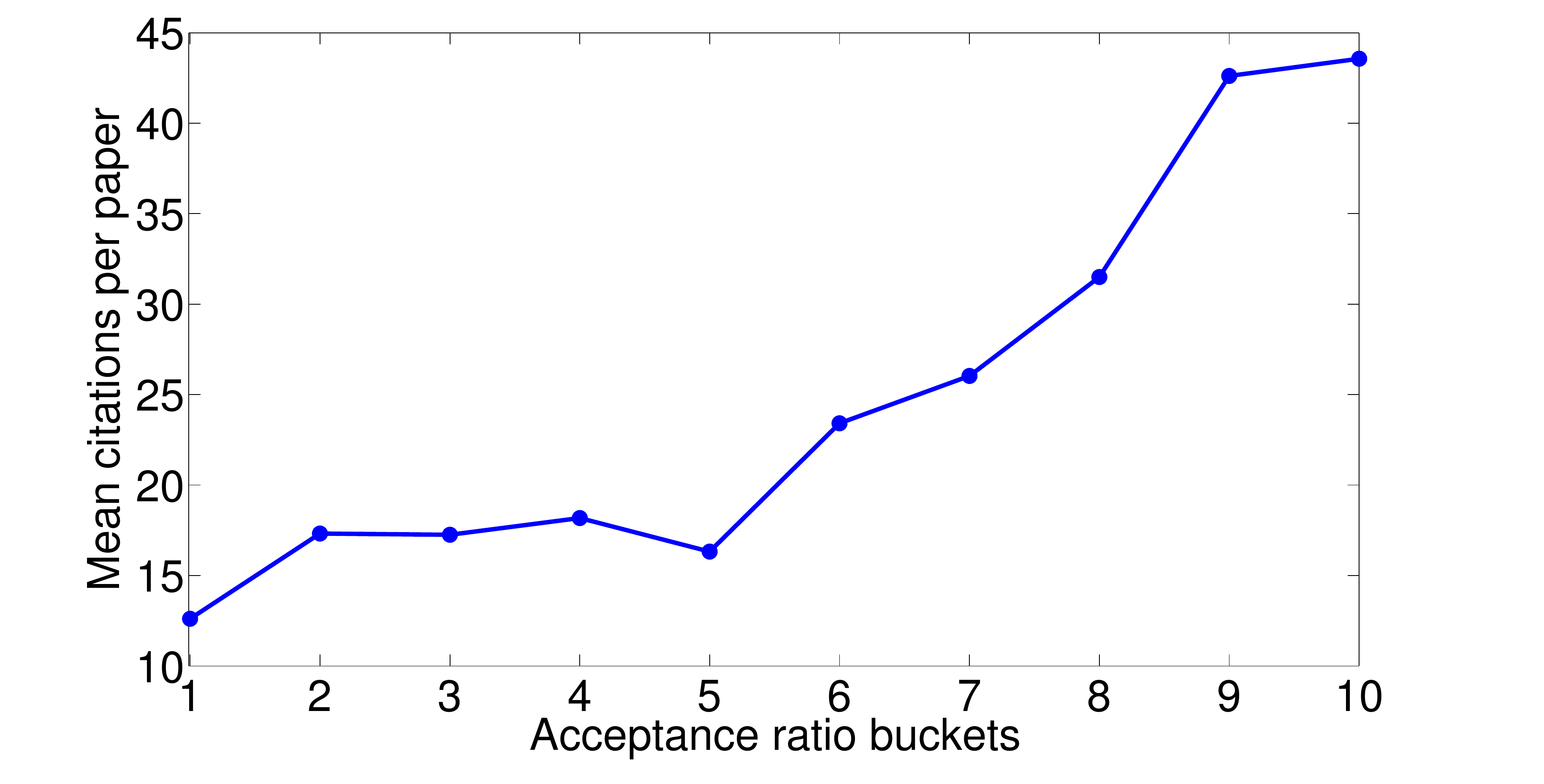} & \includegraphics[scale=0.15]{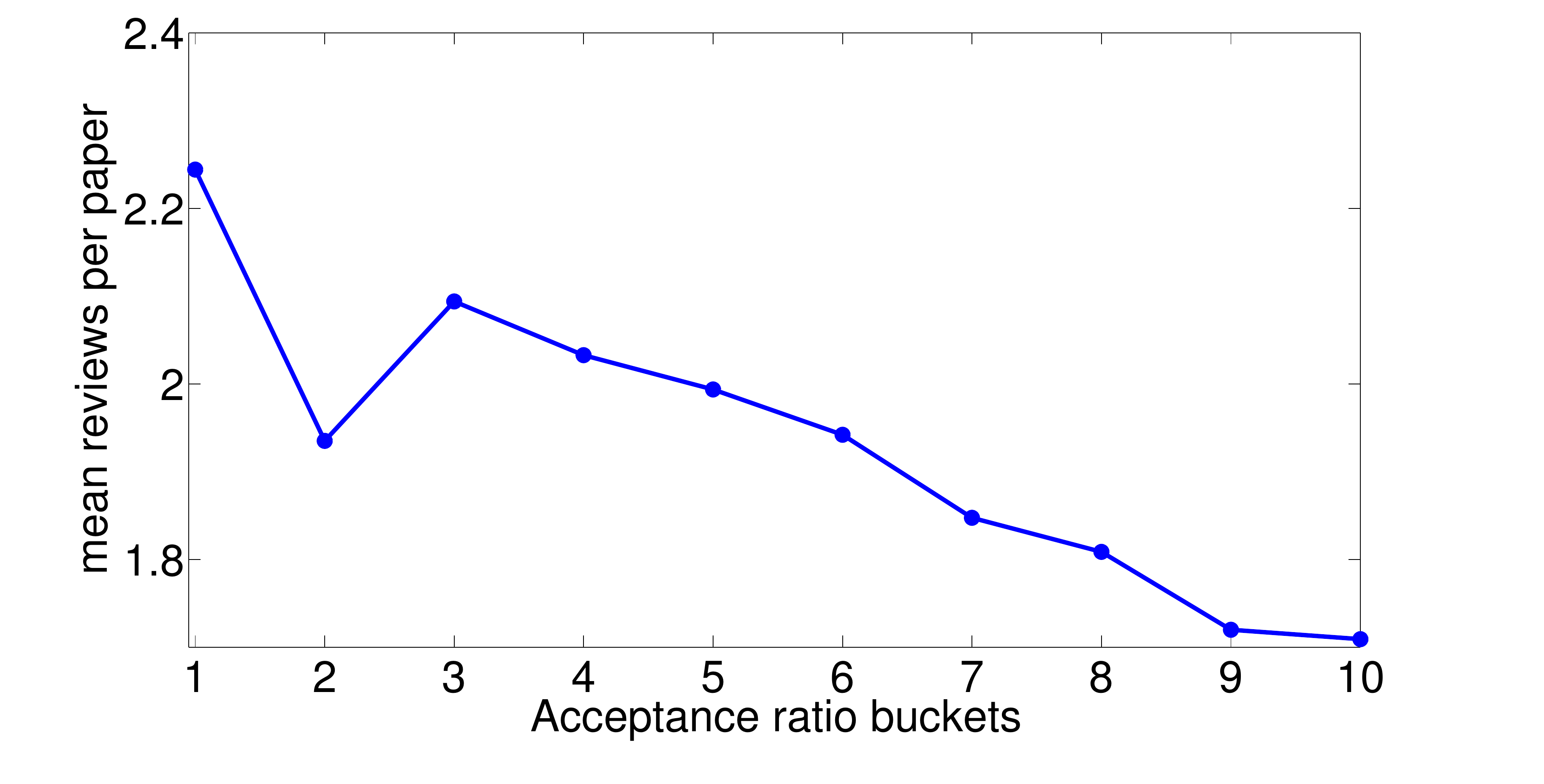} & \includegraphics[scale=0.15]{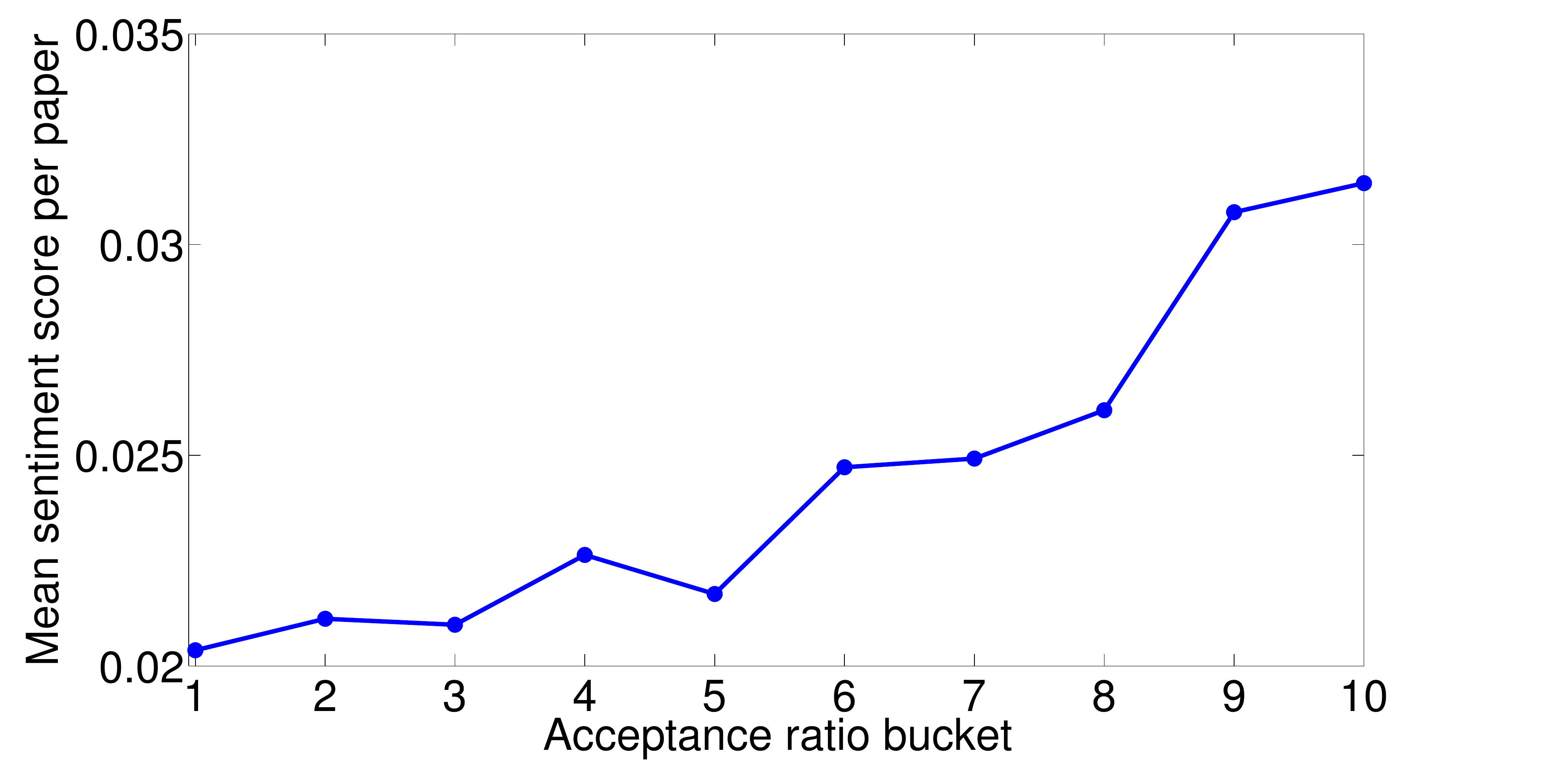}
\end{tabular}
\caption{{\bf (Left)} Mean number of citations per paper versus acceptance ratio. {\bf (Middle)} Mean number of reviews per paper versus acceptance ratio. {\bf (Right)} Mean sentiment score per paper versus acceptance ratio. Note that in each case we use acceptance ratio buckets where buckets correspond to acceptance ratio ($\geq 0.1$ and $< 0.2$), ($\geq 0.2$ and $<0.3$) and so on.}
\label{fig13}
\end{figure*}
\subsubsection{Author reputation (AR)} We analyze whether there are some specific authors whose papers always get accepted and similarly there are others whose papers always get rejected.  
For each author we define a metric called {\bf acceptance ratio} which is the fraction of submitted papers accepted in JHEP. Formally, acceptance ratio of an author $i$ is defined by: $acceptance\,ratio_{i}=\frac{accept_{i}}{accept_{i} + reject_{i}}$.    
where $accept_{i}$ and $reject_{i}$ represents respectively the number of accepted and rejected papers of the author $i$ in JHEP. We use this metric as a proxy for author reputation.
We observe that mean acceptance ratio across all the authors is $0.56$. In fact, for almost 7\% of the authors, the acceptance ratio is 1. Next we check whether the authors with high acceptance ratio have higher citations per paper. To this aim we segregate authors based on the acceptance ratio and calculate the mean number of citations per paper for these authors (refer to fig.~\ref{fig13}{\bf (Left)}). We observe an increasing trend suggesting that the authors with higher acceptance ratio tend to have higher citations. 
\begin{figure}
\centering
\includegraphics[scale=0.25]{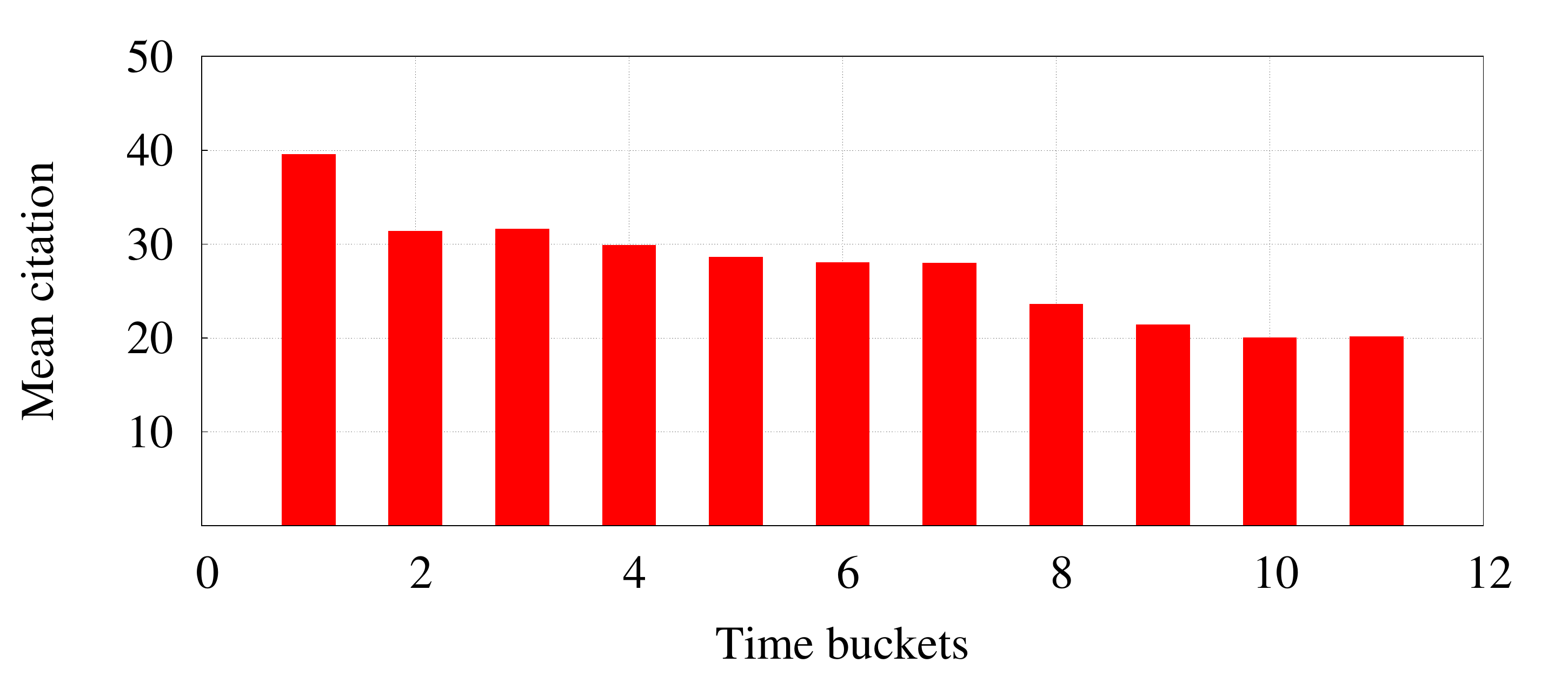}
\caption{Mean citation of the papers versus the average time(in days) between two submission. Note that  we use time buckets where buckets correspond to $<100$,($\geq 100$ and $< 200$), ($\geq 200$ and $<300$) and so on.\vspace{-2mm}}
\label{fig:prod}
\end{figure}

To check whether the authors having higher acceptance ratio are also reviewed less, we again segregate the authors based on the acceptance ratio and calculate the average number of reviews received per paper for these authors (refer to fig.~\ref{fig13}{\bf (Middle)}). We observe a decreasing trend implying that papers of authors with higher acceptance ratio are indeed reviewed less. Although there are authors with high acceptance ratio whose papers are reviewed less, they are often highly cited indicating the overall effectiveness of the review process. We further study the sentiment score of the review reports for authors with different acceptance ratios. For authors in a given  acceptance ratio bucket we calculate the average sentiment score of the review reports of their papers.
We observe that the authors having higher acceptance ratios tend to have more positive reviews on average compared to the others with lower acceptance ratios which is indicated by the increasing trend in the curve in fig.~\ref{fig13}{\bf (Right)}. 

\subsubsection{Author productivity (AP)} It is established in the literature \cite{yan2012better} that the more papers an author publishes, more are his chances of getting cited. We hence use it as a feature in predicting the long-term citation of the paper. We calculate for each author the mean time ($s_t$) between two submissions. We use $s_t$ as proxy for author productivity as low $s_t$ would indicate higher productivity rate and vice versa.  
The papers are segregated based on the corresponding author's $s_t$ and then the mean citation is calculated. Each bucket correspond to $<100$,($\geq 100$ and $< 200$), ($\geq 200$ and $<300$) and so on. We observe that more frequent the submission, more is the chance of getting citation (refer to figure \ref{fig:prod}).


\begin{table*}[]
\centering
\caption{The F-statistics value for all the features used for predicting the long-term citation of the paper.}
\label{tab:f_score}
\begin{tabular}{|l|l|l|l|l|l|l|l|l|l|l|l|l|l|l|}
\hline
Feature      & Deg  & BC    & CC    & Clus  & PR    & RR   & TS   & RL   & SNT  & AR    & AP    & RAC  & TA   & DR   \\ \hline
F-statistics & {\bf 26.1} & {\bf 29.21} & {\bf 27.72} & {\bf 17.82} & {\bf 23.34} & 6.17 & {\bf 25.6} & 14.1 & 0.94 & {\bf 18.52} & {\bf 16.49} & 3.49 & 8.68 & 7.59 \\ \hline
\end{tabular}
\end{table*}

\subsection{Reviewer based features}
\label{reviewer_analysis}

The success of the peer-review process is immensely dependent on the reviewers as they determine the quality of a paper and, consequently, the quality of the journal. We hence investigate certain reviewer behaviors (pointed out in~\cite{sikdar2016anomalies}) that could  be indicative of his/her performance.

\subsubsection{Accept ratio (RAC)} For each reviewer we define a metric called accept ratio (this is different from the acceptance ratio defined in the previous section) Formally, the {\bf accept ratio} for reviewer $j$ is $accept\,ratio_{j}=\frac{accept_{j}}{accept_{j} + reject_{j}}$ 
where $accept_{j}$ and $reject_{j}$ respectively represent the number of papers reviewer $j$ accepted and rejected. The mean accept ratio across all the reviewers is $0.62$. An accept ratio of 1 for a reviewer would mean that he accepted all the papers that were assigned to him.

\begin{figure*}
\centering
\begin{tabular}{ccc}
\includegraphics[scale=0.15]{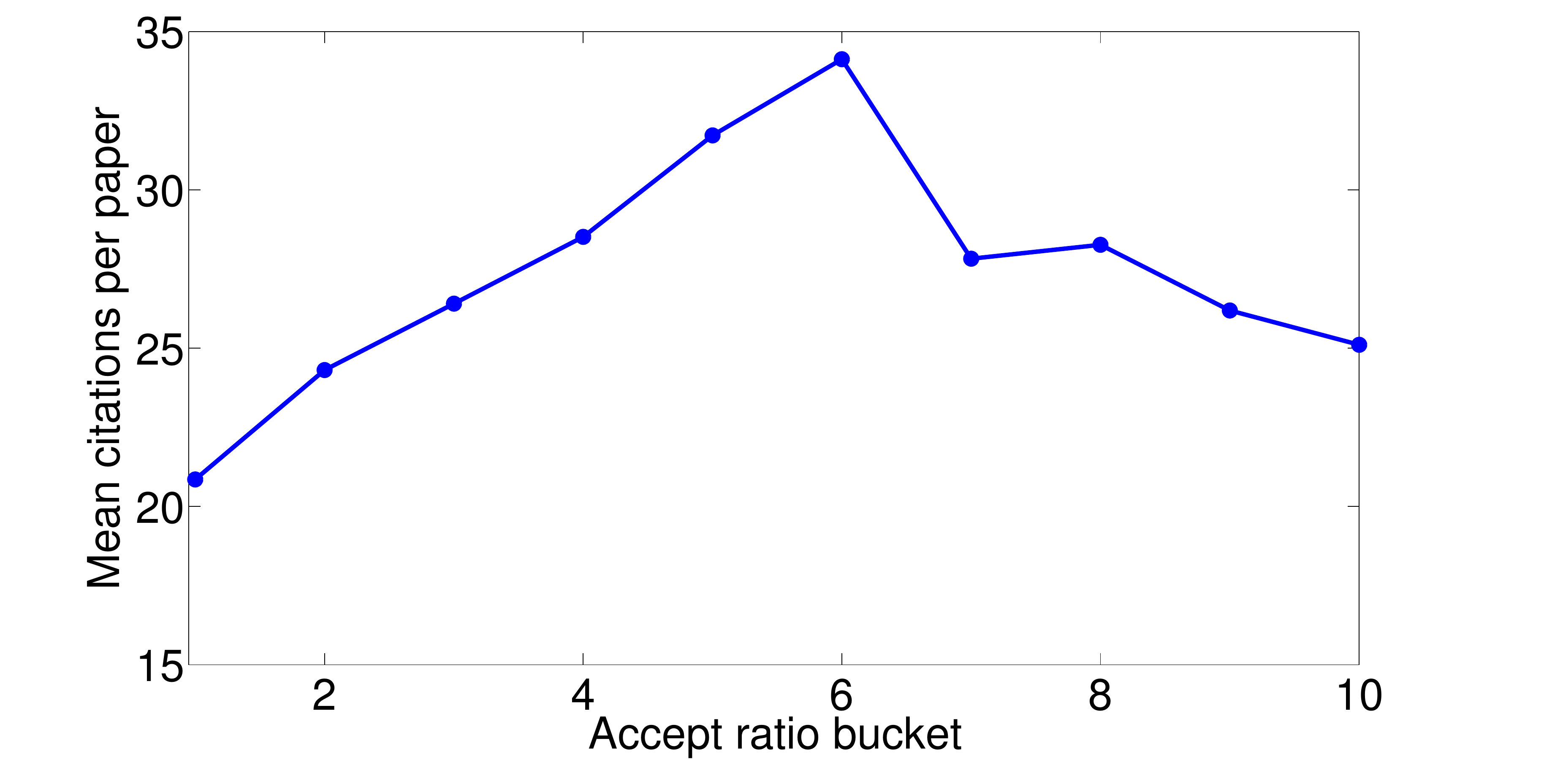} & \includegraphics[scale=0.15]{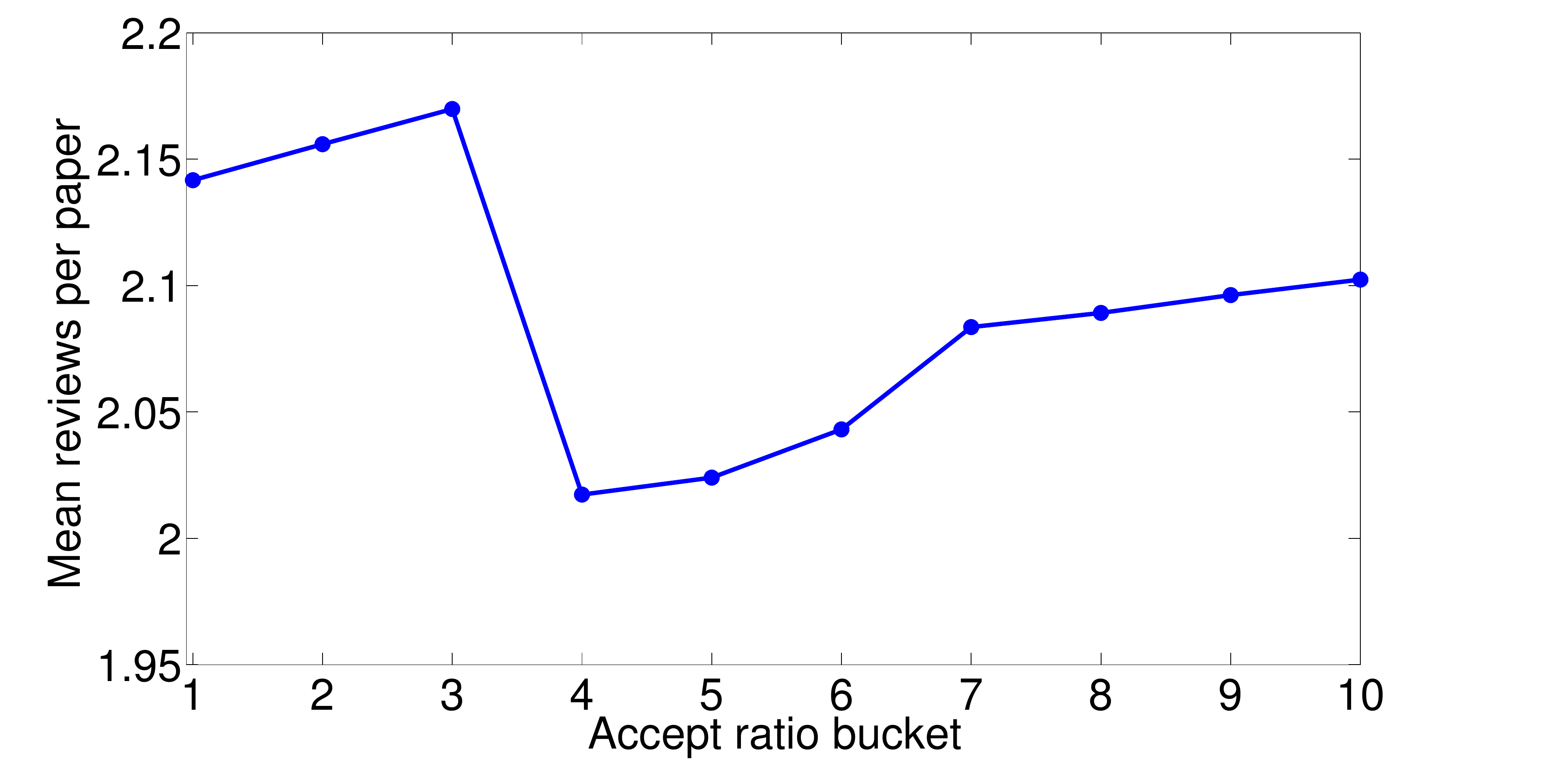} & \includegraphics[scale=0.15]{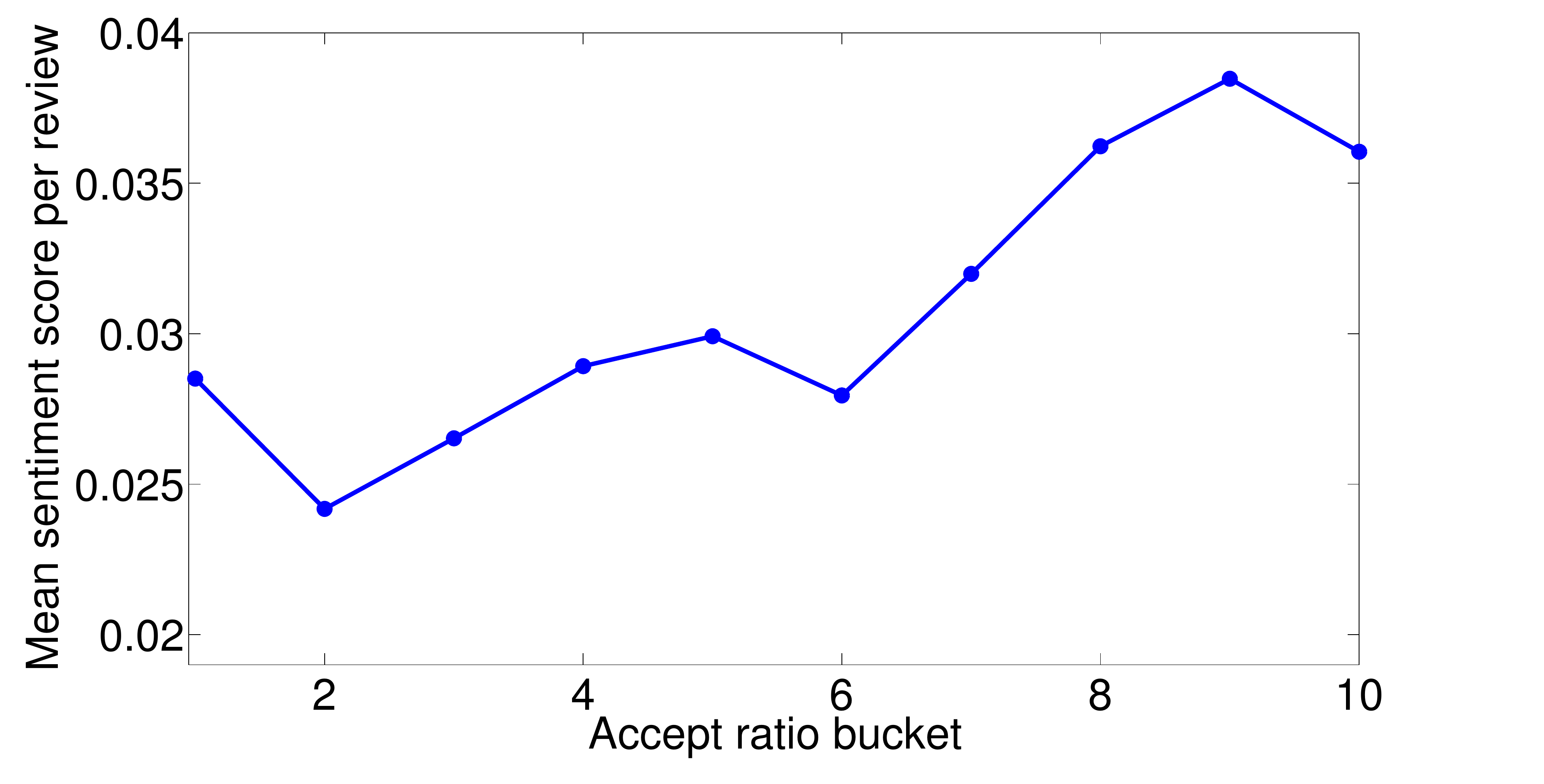}
\end{tabular}
\caption{{\bf (Left)} Mean number of citations per paper versus accept ratio. {\bf (Middle)} Mean number of reviews per paper versus accept ratio. {\bf (Right)} Mean sentiment score per paper versus accept ratio. Note that in each case we use accept ratio buckets where the buckets correspond to accept ratio ($\geq 0.1$ and $< 0.2$), ($\geq 0.2$ and $<0.3$) and so on.}
\label{fig16}
\end{figure*}

We start by investigating how well the reviewers were able to anticipate the quality of the paper. To this aim we segregate papers based on their assigned reviewer's accept ratio and calculate the mean number of citations received. In fig.~\ref{fig16}{\bf (Left)} we plot the number of citations per paper given the accept ratio of the assigned reviewer. We observe that most cited papers were reviewed by reviewers with accept ratio between 0.6 and 0.7. Surprisingly, the papers reviewed by reviewers with very low and very high accept ratio tend to garner less citations. This indicates that some papers could have been accepted just because the assigned reviewer is oriented to accept most of the papers. In fact, manual inspection indicates that many of the rejected papers that garnered large number of citations later on were mostly reviewed by reviewers with low accept ratio. We further investigate the number of reviews a reviewer with a given accept ratio suggests before he accepts or rejects a paper. For this we again segregate the papers based on the assigned reviewers accept ratio and calculate the number of rounds of reviews, these papers received on average. We present the results in fig.~\ref{fig16}{\bf (Middle)}. Since in most of the cases the same reviewer is assigned in each round of review, we observe that reviewers with a low accept ratio tend to recommend more rounds of reviews while those with higher accept ratio tend to suggest lesser number of review rounds. This indicates that the reviewers with low accept ratio often fail to improve the quality of the paper as is evident from the mean number of citations these papers receive albeit dragging the paper through multiple rounds of reviews.

To complete the analysis we also investigate the average sentiment score of the papers based on the accept ratio of the assigned reviewers. In fig.~\ref{fig16}{\bf (Right)} we plot the average sentiment score of the papers with similar accept ratio of the assigned reviewer. We observe an increasing trend indicating that the reviewers with very high accept ratio always tend to give more positive reviews as compared to others with lower accept ratio. 

\begin{figure}
\centering
\includegraphics[scale = 0.28]{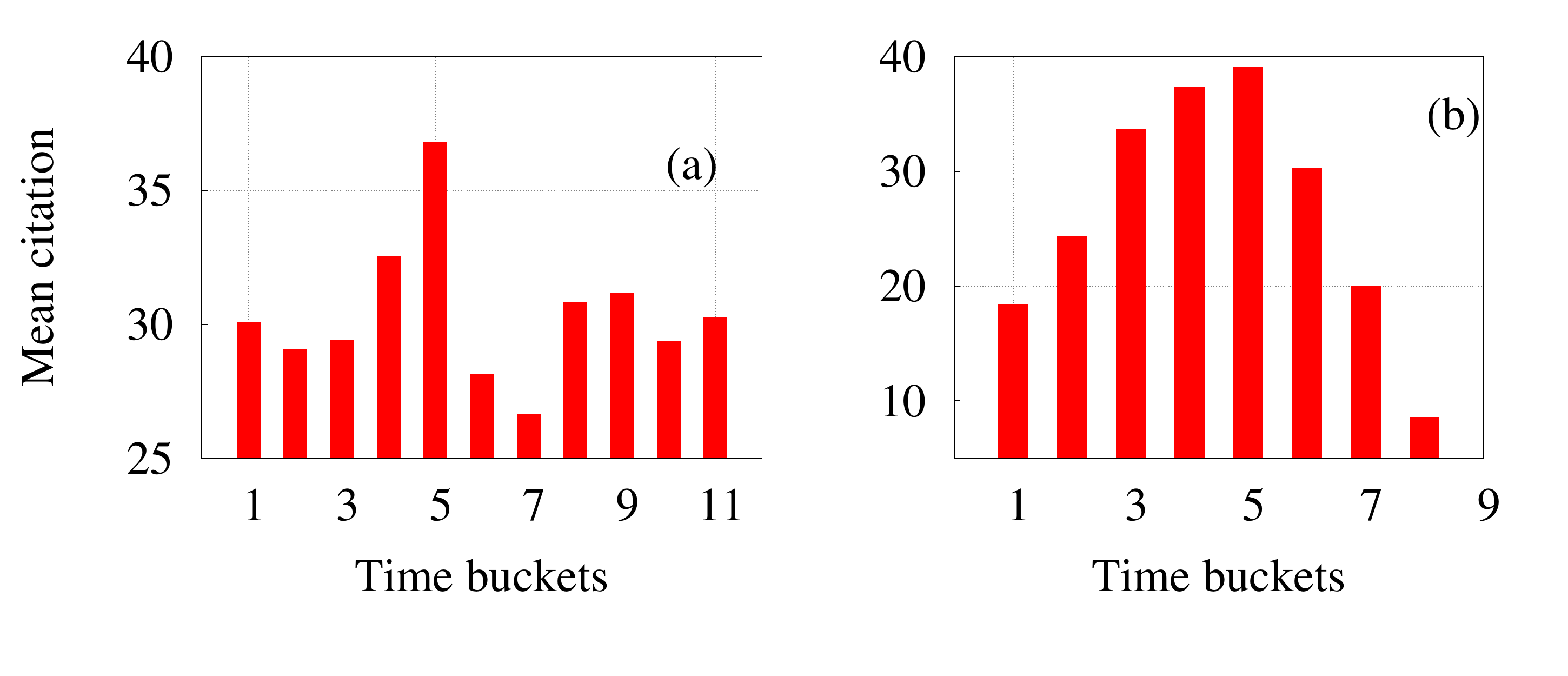}
\caption{\label{fig:rev} Mean citation of the papers versus (a) time since the last assignment for the assigned reviewer and (b) time taken by the reviewer to send the report. Note that for both the cases the times are divided into equi-sized buckets. For (a) bucket sizes are 100 each while for (b) it is 25.
}
\end{figure}

\subsubsection{Time since last assignment (TA)}

In lines of~\cite{sikdar2016anomalies} we consider the time (in number of days) since the last assignment for the assigned reviewer as an indicator of reviewer's performance and hence an indicator of the long-term citation of the paper. To verify our hypothesis we segregate the papers based on the assigned reviewer's time till last assignment and calculate the mean citation. The times are bucketed with bucket size typically $< 100$, ($\geq 100, < 200$) and so on. We observe in fig.~\ref{fig:rev}(a) that there does exist an optimal time for which the citation of the accepted paper is maximum. Further if the time since last assignment is too low or too high the long-term citation is low.   

\subsubsection{Delay in submitting the report (DR)}

We further check whether the time taken by the assigned reviewer in submitting the report is also as an indicator for his performance and hence that of the long-term citation of the paper. To this aim we calculate for each paper the time between assigned reviewer acknowledging to review the paper and the reviewer sending back the report. The papers are segregated based on this time and then the mean citation is calculated. 
The times were binned with typical bucket sizes being $>25$, $(\geq 25, < 50)$ and so on. We observe from fig.~\ref{fig:rev}(b) that the citation is maximum when the reviewer sent back the report between 50 and 75 days. The citations are comparatively less if the time is too high as well as too low.

\section{Determining the fate of the paper}
\label{performance_measure}

In this section we design a regression model to calculate the long-term citation impact of a paper. 
We perform our prediction for papers that were accepted in JHEP between 2007 and 2012. Note that for each paper, its citation till 2015 is available. Hence papers published in 2004 would have a higher citation on average compared to papers which were published in 2010 (say) due to higher exposure time. Thus, instead of calculating  the exact citation value we predict the citation rank (for each year we rank the papers based on the citations they have accrued till 2015). Further note that the papers are sorted based on the date of submission and given a paper we construct the reviewer-reviewer interaction network until its submission date (excluding). This ensures that there is no data leakage. Similarly for supporting features like acceptance ratio of an author we consider information only up to his/her last submission. 

\noindent{\bf Network features only:} Considering only the network features, we obtain the best result using support vector regression (RBF kernel) with parameters $C=100$ and $\gamma=0.01$. We perform a 10-fold cross-validation and obtain a high $R^2$ of {\bf 0.79} and a low $RMSE$ of {\bf 0.496}. 

\noindent{\bf Network + supporting features:} Considering both the network and the supporting features we obtain a further overall improvement. In specific, using support vector regression (RBF kernel) we obtain a high $R^2$ of {\bf 0.81} and a low $RMSE$ of {\bf 0.46}. The parameters were set as parameters $C=100$ and $\gamma=0.02$. We further calculate the $F$-Statistic values for all the features used in the regression task (refer to table~\ref{tab:f_score}) and observe that the network features, are in general, are more suited to the task of prediction.

Thus our system is correctly able to predict the citation rank of the paper. We believe our system could be useful in assisting the editors in deciding whether to accept or reject the papers especially in cases where the reports are contradictory.

\section{Irregular cases}

\begin{figure}
\centering
\includegraphics[scale=0.17]{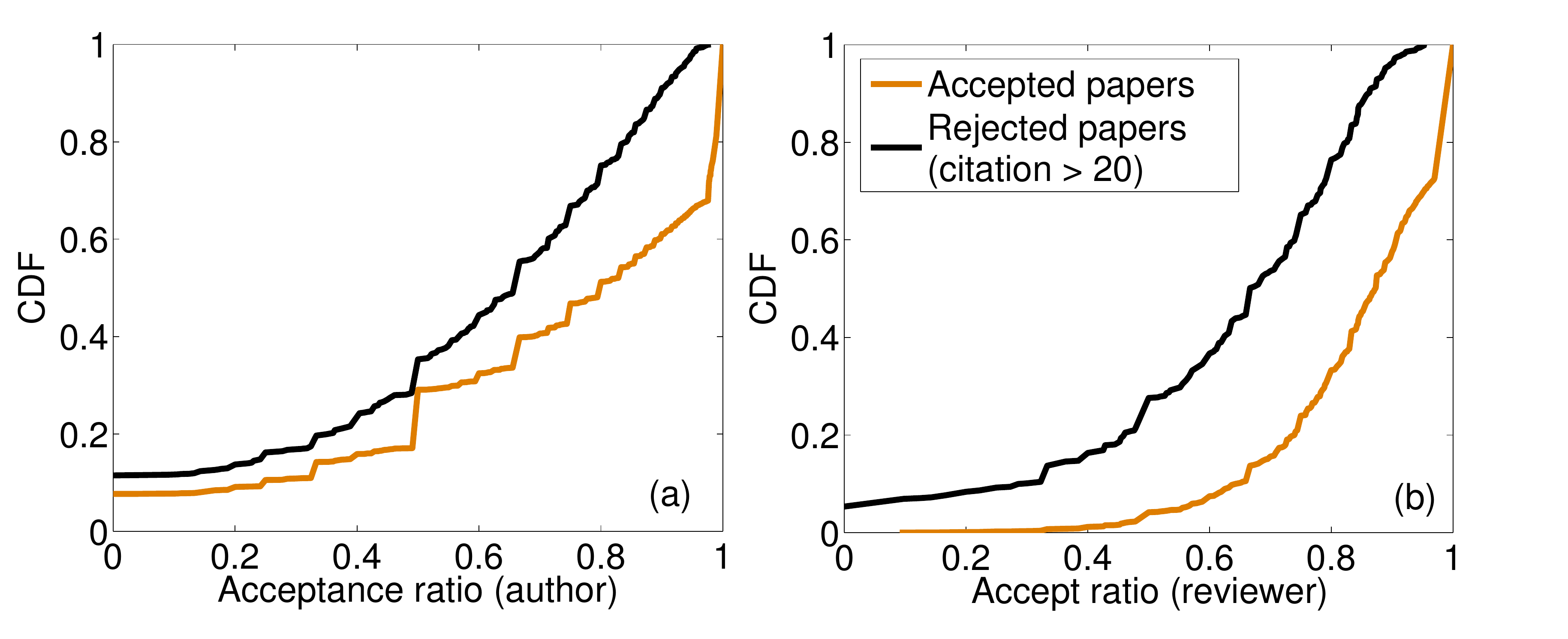}
\caption{CDF of (a) acceptance ratio of authors and (b) accept ratio of reviewers for accepted papers and highly cited rejected papers.\vspace{-5mm}}
\label{fig10}
\end{figure}

\begin{figure}
\centering
\includegraphics[scale=0.17]{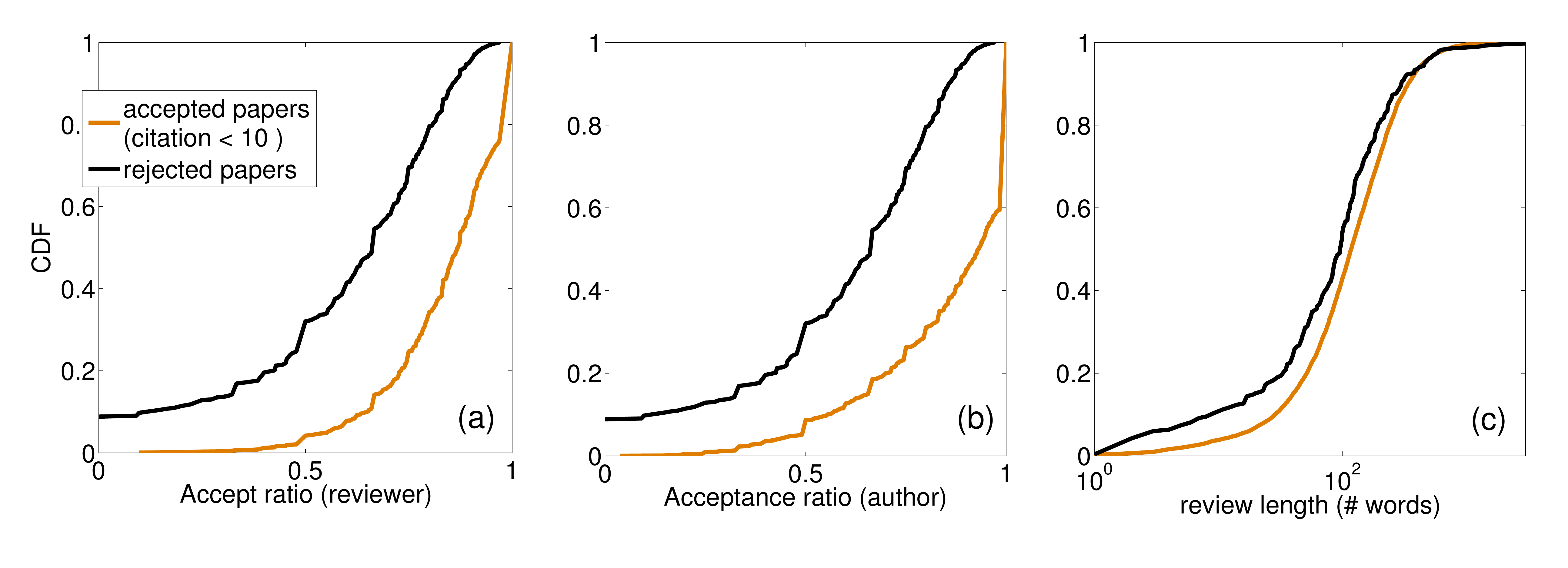}
\caption{CDF of (a) accept ratio of reviewers, (b) acceptance ratio of authors (c) length of the review text (\# words) for rejected papers and low cited accepted papers.\vspace{-5mm}}
\label{fig11}
\end{figure}

\label{irregular_cases}

In this section we investigate in more detail the irregular cases, i.e., the highly cited rejected papers and the low cited accepted papers. Note that we only consider papers which were published before 2012 so that each paper gets at least three years of exposure to the scientific community for garnering citations.

\noindent{\em Highly cited rejected papers:} We previously observed that on average the accepted papers tend to be cited more often than the rejected ones. Nevertheless, we find several papers (call it the set $P$) which were rejected at JHEP but were able to acquire more citations after getting accepted elsewhere. We consider only those papers in $P$ that have at least 20 citations which is twice the average citation of the rejected papers.
Manually looking into some of the review text we observed that in several cases the reviewer found the topic of research to be interesting but out of JHEP's scope. On deeper investigation we found that acceptance ratio of the authors of these papers is lower than that of the authors of accepted papers (fig.~\ref{fig10}(a)) indicating that author's reputation might have played a role in the rejection. We further observe that the accept ratio of the reviewers, these papers were assigned to, to be significantly less than that of the accepted papers (fig.~\ref{fig10}(b)). This indicates that the papers got assigned to stricter reviewers and hence the rejection.

\noindent{\em Low cited accepted papers:} We now look into the complementary i.e., the papers which were accepted but failed to make impact on the scientific community and accrued very low citations (typically $< 10$). Ideally, these papers should have been rejected, hence we investigate how different these are in terms of author's acceptance ratio and reviewer's accept ratio. While the authors of these papers have higher acceptance ratio (fig.~\ref{fig11}(a)), they were also assigned to reviewers who are less strict (fig.~\ref{fig11}(b)). These observations indicate that either the contributing author's reputation might have played a role in their acceptance or were lucky to have been reviewed by lenient referees. Review report also seem to be sloppy in many cases with the reviewer not even mentioning the reason for acceptance. We also observe the length of the review report (in terms of the number of words) on average to be less than that of the rejected papers (fig.~\ref{fig11}(c)). 


\section{Publisher's views}
\label{implication}
We further requested the JHEP journal administrators to survey our findings. The publishers pointed out the following observations to be of great significance -- 
(i) that reviewers excessively accepting/rejecting often fail to judge correctly the quality of the article, could be useful in assigning referees;
(ii) the number of review request does not necessarily improve the quality of the article. This observation could help in improving the efficiency of the peer-review process since both cost and time are involved for each round of review request; 
(iii) since only $10\%$ of the submissions were assigned to multiple reviewers and in most cases they failed to reach consensus, the publishers felt the need to investigate this issue more deeply by having frequent multiple assignments henceforth; 
(iv) the framework for predicting the long-term impact of the paper would be extremely helpful in assisting the editors in taking decisions. This might also aid in tracking the performance of the referees;
(v) that a significant fraction of authors have high acceptance rate at JHEP indicates a presence of at least a weak bias in the peer-review process and hence needs to be investigated;

\section{Conclusion}
\label{conclusion}

In this paper, we provided a framework for predicting the long-term citation of the paper which can be extremely helpful in assisting the editors in deciding whether to accept, reject or opt for a third opinion. We demonstrated that very simple positional properties extracted from the reviewer-reviewer interaction network are exceedingly important in determining the long-term citation of the paper. In specific, if we plug in these features in a regression model, we obtain $R^2$ = \textbf{0.79} and $RMSE$ = \textbf{0.496} in predicting the long-term citation of a paper. In addition, we also introduce a set of supporting features, based on the various properties of the paper, the authors and the assigned referees which further improved the prediction ($R^2$ = \textbf{0.81} and $RMSE$ = \textbf{0.46}). 

In the process of designing these features, we also made some key observations which are summarized below - \\ 
(i) the papers which went through lesser number of review rounds tend to be cited more on an average while the papers that were accepted after going through higher number of rounds are cited less on an average (although exceptions exist for both cases); 
(ii) although the reviewers tend to avoid highly polar words (negative or positive) in their review reports, the overall sentiment in the reports of accepted papers is more positive whereas the same is more negative for the rejected papers; 
(iii) the authors with higher acceptance ratio tend to be cited more on an average compared to those with lower acceptance ratio 
(iv) reviewers excessively accepting or rejecting most of the assigned papers often fail to correctly judge the quality of the paper;
(v) deeper investigation of the irregular cases revealed that the reputation of the author is often influential in the acceptance or rejection of the paper;
Apart from being a large-scale study that attempts to provide quantitative evidences supporting its necessity, ours is the first work that proposes definitive ways of improving the effectiveness of the scientific peer-review system.

\section{Acknowledgements}
The authors would like to thank the publication team of
Journal of High Energy Physics (JHEP) for providing 
the necessary data and they were the only ones willing to
provide it.

\end{document}